\documentclass[a4paper,UKenglish,cleveref, autoref, thm-restate]{lipics-v2021-oneside}

\usepackage{quiver}
\usepackage{tikzit}

\tikzstyle{new style 0}=[fill=white, draw=black, shape=rectangle, tikzit shape=rectangle, minimum width=0.4cm, minimum height=0.4cm]
\tikzstyle{smallrect}=[fill=white, draw=black, shape=rectangle, minimum width=0.5cm, minimum height=0.7cm]

\tikzstyle{new edge style 0}=[{-|}]
\tikzstyle{new edge style 1}=[{|-}]

\usepackage{tikz}

\usepackage{stmaryrd}

\usetikzlibrary{shapes.geometric, arrows}
\usetikzlibrary{arrows,backgrounds,shapes.geometric,shapes.misc,matrix,arrows.meta,decorations.markings}

\pgfdeclarelayer{background}
\pgfdeclarelayer{edgelayer}
\pgfdeclarelayer{nodelayer}
\pgfsetlayers{background,edgelayer,nodelayer,main}
\tikzset{baseline=-0.5ex}
\definecolor{light-gray}{gray}{.7}
\tikzstyle{none}=[inner sep=0pt]
\tikzstyle{plain}=[inner sep=0pt]
\tikzstyle{black}=[circle, draw=black, fill=black, inner sep=0pt, minimum size=3.5pt]
\tikzstyle{black-faded}=[circle, draw=light-gray, fill=light-gray, inner sep=0pt, minimum size=4pt]
\tikzstyle{white}=[circle, draw=black, fill=white, inner sep=0pt, minimum size=3.5pt]
\tikzstyle{white-faded}=[circle, draw=light-gray, fill=white, inner sep=0pt, minimum size=4.5pt]
\tikzstyle{delay}=[fill=black, regular polygon, regular polygon sides=3,rotate=-90, scale=.55]
\tikzstyle{delay-op}=[fill=black, regular polygon, regular polygon sides=3,rotate=90, scale=.55]
\tikzstyle{reg}=[draw, fill=white, rounded rectangle, rounded rectangle left arc=none, minimum height=1em, minimum width=1em, node font={\scriptsize}]
\tikzstyle{coreg}=[draw, fill=white, rounded rectangle, rounded rectangle right arc=none, minimum height=1em, minimum width=1em, node font={\scriptsize}]
\tikzstyle{rn}=[circle, draw=red, fill=red, inner sep=0pt, minimum size=4pt]
\tikzstyle{place}=[circle, draw=black, fill=white, inner sep=0pt, minimum size=8pt]
\tikzstyle{box} = [rectangle, minimum width=1cm, minimum height=1cm,text centered, draw=black, thick, fill=none]

\newcommand{\comp}{\mathop{\fatsemi}}

\makeatletter
\renewcommand{\boxed}[1]{\text{\fboxsep=.2em\fcolorbox{black}{white}{\m@th$\displaystyle#1$}}}
\makeatother

\usetikzlibrary{calc}
\def\len{0.25}

\tikzset{
  wire/.style={rounded corners=3, line width=.5pt}
, iden/.style={line width=.5pt}
, dot/.style={draw,fill=black, circle, inner sep=1pt}
, wdot/.style={draw,fill=white, circle, inner sep=1pt}
, edot/.style={draw,fill=red!40, circle, inner sep=1pt}
, cdot/.style={draw,fill=#1, circle, inner sep=1pt}
, labeled/.style={draw,fill=white, inner sep=1pt}
}

\def\akasa{
\draw[densely dotted] (0,0) rectangle (1,1);
}

\def\braid{
  \draw[draw=none] (0,0) rectangle (1,1);
\draw[wire] (0,\len) -- (\len,\len) -- (3*\len, 3*\len) -- (1,3*\len);
\draw[white,wire] (0,3*\len) -- (\len,3*\len) -- (3*\len, \len) -- (1,\len);
}
\def\id{
  \draw[draw=none] (0,0) rectangle (1,1);
\draw[iden] (0,2*\len) -- (1,2*\len);
}
\def\twoid{
\draw[iden] (0,\len) -- (1,\len);
\draw[iden] (0,3*\len) -- (1,3*\len);
}
\def\mult{
  \draw[draw=none] (0,0) rectangle (1,1);
\draw[wire] (0,\len) -| (2*\len,3*\len) -- (0,3*\len);
\draw[wire] (2*\len,2*\len) -- (1,2*\len);
\node[dot] at (2*\len,2*\len) {};
}

\def\comult{
  \draw[draw=none] (0,0) rectangle (1,1);
\draw[wire] (4*\len,\len) -| (2*\len,3*\len) -- (1,3*\len);
\draw[wire] (2*\len,2*\len) -- (0,2*\len);
\node[dot] at (2*\len,2*\len) {};
}

\def\emult{
\draw[wire] (0,\len) -| (2*\len,3*\len) -- (0,3*\len);
\draw[wire] (2*\len,2*\len) -- (1,2*\len);
\node[edot] at (2*\len,2*\len) {};
}

\def\unit{
  \draw[draw=none] (0,0) rectangle (1,1);
\draw[wire] (2*\len,2*\len) -- (1,2*\len);
\node[dot] at (2*\len,2*\len) {};
}
\def\eunit{
\draw[wire] (2*\len,2*\len) -- (1,2*\len);
\node[edot] at (2*\len,2*\len) {};
}
\def\lid{
  \draw[draw=none] (0,0) rectangle (1,1);
\draw[iden] (0,\len) -- (1,\len);
}
\def\uid{
  \draw[draw=none] (0,0) rectangle (1,1);
\draw[iden] (0,3*\len) -- (1,3*\len);
}
\def\mor#1{
  \draw[draw=none] (0,0) rectangle (1,1);
\id
\node[draw, fill=white, inner sep=1.5pt] at (2*\len,2*\len) {\tiny $#1$};
}
\def\lmor#1{
\lid
\node[draw, fill=white, inner sep=1.5pt] at (2*\len,\len) {\tiny $#1$};
}
\def\umor#1{
\uid
\node[draw, fill=white, inner sep=1.5pt] at (2*\len,3*\len) {\tiny $#1$};
}
\def\dcouni#1{
  \draw[draw=none] (0,0) rectangle (1,1);
\draw[wire] (2*\len,2*\len) -- ++(2*\len,0);
\node[draw, fill=white, inner sep=1.5pt] at (2*\len,2*\len) {\tiny $#1$};
}

\def\lcomult{
\draw[wire] (4*\len,0) -| (2*\len,2*\len) -- (1,2*\len);
\draw[wire] (2*\len,\len) -- (0,\len);
\node[dot] at (2*\len,\len) {};
}

\def\ucomult{
\draw[wire] (4*\len,2*\len) -| (2*\len,1) -- (1,1);
\draw[wire] (2*\len,3*\len) -- (0,3*\len);
\node[dot] at (2*\len,3*\len) {};
}

\def\counit{
\draw[wire] (0,2*\len) -- (2*\len,2*\len);
\node[dot] at (2*\len,2*\len) {};
}

\def\ucounit{
\draw[wire] (0,3*\len) -- (2*\len,3*\len);
\node[dot] at (2*\len,3*\len) {};
}
\def\lcounit{
\draw[wire] (0,\len) -- (2*\len,\len);
\node[dot] at (2*\len,\len) {};
}
\def\thingy#1{
  \draw[wire] (4*\len,\len) -| (2*\len,3*\len) -- (1,3*\len);
  \draw[wire] (2*\len,2*\len) -- (0,2*\len);
  \draw[fill=white] (\len,\len-.1) rectangle (3*\len,3*\len+.1) node[pos=.5] {$\scriptscriptstyle #1$};
}

\def\turn{
	\draw[rounded corners=3, line width=.5pt] (0,\len) -- (\len,\len) -- (3*\len, 3*\len) -- (4*\len,3*\len);
}
\def\coturn{
	\draw[rounded corners=3, line width=.5pt] (0,3*\len) -- (\len,3*\len) -- (3*\len, \len) -- (4*\len,\len);
}

\def\tenmult{
\draw[wire] (0,\len) -| (2*\len,3*\len) -- (0,3*\len);
\draw[wire] (2*\len,2*\len) -- (1,2*\len);
\node[fill=white,inner sep=-1.25pt,circle] at (2*\len,2*\len) { $\otimes$};
}

\newcommand{\ezs}[2]{
\begin{scope}[#2]
#1
\end{scope}
}

\newcommand{\step}[2][1]{\ezs{#2}{xshift=#1 cm}}

\newcommand{\leftLabel}[2][0]{
\node[left] at (#1,2*\len) {$#2 =\quad$};
}
\newcommand{\rightLabel}[2]{
\node[right] at (#2,2*\len) {$\quad = #1$};
}

\newcommand{\umult}[1][black]{
\ezs{
\draw[wire] (0,\len) -| (2*\len,3*\len) -- (0,3*\len);
\draw[wire] (2*\len,2*\len) -- (1,2*\len);
\node[cdot=#1] at (2*\len,2*\len) {};
}{yshift=\len cm}
}

\newcommand{\lmult}[1][black]{
\ezs{
\draw[wire] (0,\len) -| (2*\len,3*\len) -- (0,3*\len);
\draw[wire] (2*\len,2*\len) -- (1,2*\len);
\node[cdot=#1] at (2*\len,2*\len) {};
}{yshift=-\len cm}
}

\newcommand*{\Scale}[2][4]{\scalebox{#1}{\ensuremath{#2}}}%

\def\cothingy#1{
  \ezs{\thingy{#1}}{xscale=-1}
}

\def\multi#1{
\draw[wire] (0,5*\len) -| (2*\len,7*\len) -- (0,7*\len);
\draw[wire] (0,\len) -| (2*\len,3*\len) -- (0,3*\len);
\ezs{\draw[wire] (3*\len,2*\len) -- ++(\len,0);
\draw[fill=white, thin] (\len,\len-.15) rectangle (3*\len,3*\len+.15) node[pos=.5,] {$\scriptscriptstyle #1$};}{yscale=2}
}

\newcommand{\up}[2][1]{
    {\ezs{#2}{yshift=#1*\len cm}}
}
\newcommand{\down}[2][1]{
    {\ezs{#2}{yshift=-#1*\len cm}}
}

\newcommand{\tmor}[3]{
\mor{#1}
\node[left] at (0,2*\len) {\tiny $#2$};
\node[right] at (1,2*\len) {\tiny $#3$};
}

\newcommand{\gau}[1]{
\node[left] at (0,2*\len) {\tiny $#1$};
}

\newcommand{\dro}[1]{
\node[right] at (1,2*\len) {\tiny $#1$};
}

\newcommand{\eql}[1]{
\node at (#1,2*\len) {$=$};
}

\newcommand{\from}{\mathrel{:}}

\title{Regular Monoidal Languages} 

\author{Matthew Earnshaw}{Department of Software Science, Tallinn University of Technology, Estonia \and \url{https://ioc.ee/~matt} }{matthew.earnshaw@taltech.ee}{https://orcid.org/0000-0001-8236-2811}{}

\author{Pawe\l{} Soboci\'{n}ski}{Department of Software Science, Tallinn University of Technology, Estonia \and \url{https://ioc.ee/~pawel} }{pawel.sobocinski@taltech.ee}{https://orcid.org/0000-0002-7992-9685}{}

\authorrunning{M. Earnshaw and P. Soboci\'{n}ski} 

\Copyright{Matthew Earnshaw and Pawe\l{} Soboci\'{n}ski} 



\category{} 

\relatedversion{} 

\funding{This research was supported by the ESF funded Estonian IT Academy research measure (project 2014-2020.4.05.19- 0001). The second author was additionally supported by the Estonian Research Council grant PRG1210.}

\acknowledgements{We would like to thank Ed Morehouse for extensive discussions concerning this work, and Tobias Heindel for discussion of his erstwhile project that partially inspired ours.}

\nolinenumbers 

\hideLIPIcs  

\EventEditors{John Q. Open and Joan R. Access}
\EventNoEds{2}
\EventLongTitle{42nd Conference on Very Important Topics (CVIT 2016)}
\EventShortTitle{CVIT 2016}
\EventAcronym{CVIT}
\EventYear{2016}
\EventDate{December 24--27, 2016}
\EventLocation{Little Whinging, United Kingdom}
\EventLogo{}
\SeriesVolume{42}
\ArticleNo{23}

\usepackage[color=yellow,textsize=footnotesize]{todonotes}
\usepackage{mathtools}
\makeatletter
\def\slashedarrowfill@#1#2#3#4#5{%
  $\m@th\thickmuskip0mu\medmuskip\thickmuskip\thinmuskip\thickmuskip
  \relax#5#1\mkern-7mu%
  \cleaders\hbox{$#5\mkern-2mu#2\mkern-2mu$}\hfill
  \mathclap{#3}\mathclap{#2}%
  \cleaders\hbox{$#5\mkern-2mu#2\mkern-2mu$}\hfill
  \mkern-7mu#4$%
}
\def\rightslashedarrowfill@{%
  \slashedarrowfill@\relbar\relbar\mapstochar\rightarrow}
\newcommand\xslashedrightarrow[2][]{%
  \ext@arrow 0055{\rightslashedarrowfill@}{#1}{#2}}
\makeatother

\usepackage{xparse}

\NewDocumentCommand\Par{o}
{
  \IfNoValueTF{#1}
  {\textsf{Par}}
  {\textsf{Par}_{#1}}
}

\NewDocumentCommand\CRel{o}
{
  \IfNoValueTF{#1}
  {\textsf{CRel}}
  {\textsf{CRel}_{#1}}
}

\NewDocumentCommand\Lang{}
{
  \mathscr{L}
}

\NewDocumentCommand\Rel{o}
{
  \IfNoValueTF{#1}
  {\textsf{Rel}}
  {\textsf{Rel}_{#1}}
}

\NewDocumentCommand\detr{o}
{
  \IfNoValueTF{#1}
  {\mathscr{D}}
  {\mathscr{D}(#1)}
}

\NewDocumentCommand\endo{O{\mathscr{C}}m}
{
  {#1}_{#2}
}

\begin{document}

\maketitle

\begin{abstract}
  We introduce regular languages of morphisms in free monoidal categories, with their associated grammars and automata. These subsume the classical theory of regular languages of words and trees, but also open up a much wider class of languages over string diagrams. We use the algebra of monoidal and cartesian restriction categories to investigate the properties of regular monoidal languages, and provide sufficient conditions for their recognizability by deterministic monoidal automata.
\end{abstract}

\section{Introduction} \label{sec:intro}
Classical formal language theory has been extended to various kinds of algebraic structures, such as infinite words, rational sequences, trees, countable linear orders, graphs of bounded tree width, etc. In recent years, the essential unity of the field has been better understood \cite{mmso,eilenbergforfree}.
Such structures can often be seen as algebras for monads on the category of sets, and sufficient conditions exist~\cite{mmso} for formal language theory to extend to their algebras.

In this paper, we make a first step into a programme of extending language theory to higher-dimensional algebraic structures. Here we make the step from monoids to 2-monoids, better known as monoidal categories.

We introduce a categorial framework for reasoning about languages of morphisms in strict monoidal categories -- including their associated grammars and automata. We show how these include classical and tree automata, but also open up a wilder world of string diagram languages. By investigating the morphisms in monoidal categories from the perspective of language theory, this work contributes to research into the computational manipulation of string diagrams, and so their usage in industrial strength applications.

\section{Related work} \label{sec:related_work}
Bossut \cite{bossut} studied rational languages of planar acyclic graphs and proved a Kleene theorem for a class of such languages. Bossut's graph languages feature initial and final states, whereas our languages consist of scalar morphisms, which more neatly generalizes the theory of regular string and tree languages. Bossut introduces a notion of automaton for these languages, but these lack a state machine denotation -- being more similar to our grammars.

In \cite{Heindel2017AMT}, Heindel recasts Bossut's approach using monoidal categories. Unfortunately the purported Myhill-Nerode result was incorrect, due to a flawed definition of syntactic congruence. We rectify this in Section \ref{sec:algebraic}, but a Myhill-Nerode type theorem remains open.

Zamdzhiev \cite{zamdzhiev2016} introduced context-free languages of string diagrams using the string graph representation of string diagrams and the machinery of context-free graph grammars. In contrast, our approach does not require an intermediate representation of string diagrams as graphs: we work directly with morphisms in monoidal categories. This allows us to use the algebra of monoidal categories to reason about properties of monoidal languages.

Winfree et al. \cite{10.1371/journal.pbio.0020424} use DNA self-assembly to simulate cellular automata and Wang tile models of computation. The kinds of two-dimensional languages obtained in this way can be seen quite naturally as regular monoidal languages, as illustrated in Example \ref{ex:sierpinski}.

Walters' note \cite{WALTERS1989199} on regular and context-free grammars served as a starting point for our definition of regular monoidal grammar. Rosenthal \cite{Rosenthal1995}, developing some of the ideas of Walters, defined automata as relational presheaves, which is similar in spirit to our functorial definition of monoidal automata. The framework of Colcombet and Petrişan \cite{lmcs:6213} considering automata as functors is also close in spirit to our definition of monoidal automata. However, all of these papers are directed towards questions involving classical one-dimensional languages, rather than languages of diagrams as in the present paper.

  Fahrenberg et al. \cite{fahrenberg} investigated languages of higher-dimensonal automata, a well-established model of concurrency. We might expect that the investigations of the present paper correspond to a detailed study of a particular low-dimensional case of such languages, but the precise correspondence between these notions is unclear.

\section{Regular monoidal grammars and regular monoidal languages} \label{sec:mon_grammars}
A \emph{monoidal grammar} is a finite specification for the construction of string diagrams: i.e.\ morphisms in free monoidal categories (more specifically, free pros). We introduce \emph{regular monoidal grammars}, an analogue of classical (right-) regular grammars, and their equivalent representation as non-deterministic \emph{monoidal automata}. We begin by recalling the notion of monoidal graph and how they present free monoidal categories.

\subsection{Monoidal graphs and free pros}

\begin{definition}
  A monoidal graph $\mathcal{G}$ consists of sets $E_{\mathcal{G}},V_{\mathcal{G}}$ and functions $dom,cod : E_{\mathcal{G}} \rightrightarrows V_{\mathcal{G}}^{*}$ where $V_{\mathcal{G}}^{*}$ is the underlying set of the free monoid. The elements of $E_{\mathcal{G}}$ are called \emph{generators}, and for a generator $\gamma \in E_{\mathcal{G}}$, $dom(\gamma), cod(\gamma)$ are the domain, codomain (resp.) types of $\gamma$.
\end{definition}

Diagrammatically, a monoidal graph can be pictured as a collection of boxes, labelled by elements of $E_{\mathcal{G}}$ with wires entering on the left and exiting on the right, labelled by types given by the functions $dom,cod$. For example, the following depicts the monoidal graph $\mathcal{G}$ with $E_{\mathcal{G}} = \{\gamma, \gamma'\}, V_{\mathcal{G}} = \{A,B\}, dom(\gamma) = AB, cod(\gamma) = ABA, dom(\gamma') = A, cod(\gamma') = BB$:

\begin{center}
\begin{minipage}{.25\textwidth}
\begin{tikzpicture}[node distance=1cm]
  \node [style=none] (l1) at (-1.2,0.25) {A};
  \node [style=none] (l2) at (-1.2,-0.25) {B};
  \draw (-1,-0.25) -- (-0.5,-0.25);
  \draw (-1,0.25) -- (-0.5,0.25);
  \node (gamma) [box] {$\gamma$};
  \draw (0.5,0) -- (1,0);
  \draw (0.5,0.3) -- (1,0.3);
  \draw (0.5,-0.3) -- (1,-0.3);
  \node [style=none] (r1) at (1.2,0.3) {A};
  \node [style=none] (r2) at (1.2,0) {B};
  \node [style=none] (r3) at (1.2,-0.3) {A};
\end{tikzpicture}
\end{minipage} \begin{minipage}{.25\textwidth}
\hspace*{5mm}
\begin{tikzpicture}[node distance=2cm]
  \node [style=none] (r1) at (-1.2,0) {A};
  \draw (-1,0) -- (-0.5,0);
  \node (gamma') [box] {$\gamma'$};
  \draw (0.5,-0.25) -- (1,-0.25);
  \draw (0.5,0.25) -- (1,0.25);
  \node [style=none] (r1) at (1.2,0.25) {B};
  \node [style=none] (r2) at (1.2,-0.25) {B};
\end{tikzpicture}
\end{minipage}
\end{center}

Given that we are interested in finite state machines over finite alphabets, we shall work exclusively with finite monoidal graphs, i.e. those in which $E_{\mathcal{G}}$ and $V_{\mathcal{G}}$ are both finite sets.

\begin{definition}
 A morphism $\Psi \from \mathcal{G}' \to \mathcal{G}$ of monoidal graphs is a pair of functions  $V_\Psi \from V_{\mathcal{G}} \to V_{\mathcal{G}'}, E_\Psi \from E_{\mathcal{G}} \to E_{\mathcal{G}'}$ such that $dom \comp V_\Psi^{*} = E_\Psi \comp dom$ and $cod \comp V_\Psi^{*} = E_\Psi \comp cod$.
\end{definition}

Monoidal graphs and their morphisms form a category $\mathsf{MonGraph}$. Recall that a (coloured) pro is a strict monoidal category whose monoid of objects is free (on the set of ``colours''). There is a category $\mathsf{Pro}$ with objects pros and morphisms strict monoidal functors whose action on objects is determined by a function between their sets of colours. We call these \emph{pro morphisms}. (Coloured) props are pros that are also \emph{symmetric} (strict) monoidal categories.

Pros (and props) are monadic over monoidal graphs: the forgetful functor $\mathscr{U} : \mathsf{Pro} \to \mathsf{MonGraph}$ has a left adjoint $\mathscr{F} : \mathsf{MonGraph} \to \mathsf{Pro}$, and $\mathsf{Pro}$ is equivalent to the category of algebras for the induced monad on $\mathsf{MonGraph}$ (see \cite[\S 2.3]{10.1093/logcom/exx029}). $\mathscr{F}$ sends a monoidal graph $\mathcal{G}$ to a pro $\mathscr{F}\mathcal{G}$ whose set of objects is $V_{\mathcal{G}}^{*}$ and whose morphisms are \emph{string diagrams} (see Appendix \ref{appendix:string}).

\subsection{Monoidal languages and regular monoidal grammars}
Classically, a language over an alphabet $\Sigma$ is a subset of the free monoid $\Sigma^{*}$. A \emph{monoidal language} is defined similarly, replacing free monoids with free pros over a \emph{monoidal alphabet}:

\begin{definition} \label{defn:monoidal_alphabet}
   A monoidal alphabet $\Gamma$ is a finite monoidal graph where $V_{\Gamma}$ is a singleton.
\end{definition}

 For a generator $\gamma$ of a monoidal alphabet, we refer to $dom(\gamma), cod(\gamma)$ as the arity, coarity (resp.) of $\gamma$, writing $ar(\gamma)$, $coar(\gamma)$. Such generators are drawn with ``untyped'' wires.

\begin{definition} \label{defn:language}
  A monoidal language $L$ over a monoidal alphabet $\Gamma$ is a subset $L \subseteq \mathscr{F}\Gamma(0,0)$ of morphisms with arity and coarity 0 in the free pro generated by $\Gamma$.
\end{definition}

\begin{remark}
  The restriction to arity and coarity zero (i.e. \emph{scalar}) morphisms may appear arbitrary. However, we will see in Section \ref{sec:classical} that this captures and explains the classical definitions of finite-state automata over words and trees. It also leads to more concise definitions in our theory.
\end{remark}

\emph{Regular monoidal grammars} specify monoidal languages that are an analogue of classical regular languages. They can be obtained by taking Walters' \cite{WALTERS1989199} definition of regular language and replacing the adjunction between reflexive graphs and categories with that between monoidal graphs and pros. As shown in Section \ref{sec:classical}, they include the classical definitions of regular tree and word languages as grammars over monoidal alphabets of a particular shape.

\begin{definition} \label{defn:monoidal_grammar}
  A regular monoidal grammar is a morphism of finite monoidal graphs $\Psi : \mathcal{M} \to \Gamma$ where $\Gamma$ is a monoidal alphabet.
\end{definition}

Intuitively, a regular monoidal grammar is a labelling of the edges of $\mathcal{M}$ by generators in $\Gamma$. Indeed, the vertex function $V_\Psi : V_{\mathcal{M}} \to \{\bullet\}$ is unique, so the grammar is determined by its edge function $E_\Psi : E_{\mathcal{M}} \to E_\Gamma$, sending edges to their labels. In Section \ref{sec:pro_aut} we show that this data determines a transition system with states words $w \in V_{\mathcal{M}}^{*}$.

\begin{remark}
Every regular monoidal grammar determines a pro morphism between free pros, $\mathscr{F}\Psi : \mathscr{F}\mathcal{M} \to \mathscr{F}\Gamma$, which we may also refer to as a regular monoidal grammar.
\end{remark}

For any string diagram $s \in \mathscr{F}\Gamma$ over an alphabet $\Gamma$, we can think of the set of string diagrams $\mathscr{F}\Psi^{-1}(s)$ as a set of possible ``parsings'' of that diagram.

\begin{remark} \label{remark:graphical}
  We represent regular monoidal grammars diagrammatically by drawing the monoidal graph $\mathcal{M}$ as above, but labelling each box $e \in E_{\mathcal{M}}$ with $E_{\Psi}(e)$. The resulting diagram is not in general a diagram of a monoidal graph, since it may contain boxes with the same label but different domain or codomain types.  Examples are given below.
\end{remark}

\subsection{Regular monoidal languages}

A regular monoidal grammar determines a monoidal language as follows:

\begin{definition}
  Given a regular monoidal grammar $\Psi : \mathcal{M} \to \Gamma$, the image under $\mathscr{F}\Psi$ of the endo-hom-set of the monoidal unit $\varepsilon$ in $\mathscr{F}\mathcal{M}$ is a monoidal language $\mathscr{F}\Psi[\mathscr{F}\mathcal{M}(\varepsilon,\varepsilon)] \subseteq \mathscr{F}\Gamma(0,0)$.
\end{definition}

We call the class of languages determined by regular monoidal grammars the \emph{regular monoidal languages}. We shall see that they are precisely the languages accepted by \emph{non-deterministic monoidal automata} (Section \ref{sec:pro_aut}).
The basic idea is that a ``word'' is a scalar string diagram, i.e. one with no ``dangling wires''. The language of a monoidal grammar then consists of those scalar string diagrams that can be given a parsing. Parsings can be visually explained using the graphical notation for grammars (Remark \ref{remark:graphical}). A morphism in the language defined by a grammar is any string diagram that can be built using the ``typed'' building blocks, such that there are no dangling wires, and then erasing the types on the wires. The following examples of regular monoidal grammars illustrate this idea:

\begin{example}[Balanced parentheses] \label{ex:brackets}
  Recall that the Dyck language, the language of balanced parentheses, is a paradigmatic example of a non-regular word language. However, we can recognize balanced parentheses using the regular monoidal grammar shown below left. An example of a morphism in the language defined by this grammar is shown on the right.
  \begin{center}
    \begin{tikzpicture}
	\begin{pgfonlayer}{nodelayer}
		\node [style=none] (0) at (0.5, 0) {};
		\node [style=none] (1) at (0.2, 0) {};
                \node [style=none] (2) at (0.7, 0) {A};
	\end{pgfonlayer}
	\begin{pgfonlayer}{edgelayer}
		\draw [style=new edge style 1, in=180, out=0] (1.center) to (0.center);
	\end{pgfonlayer}
\end{tikzpicture}

\begin{tikzpicture}
	\begin{pgfonlayer}{nodelayer}
		\node [style=smallrect] (r) at (0, 0) {(};
		\node [style=none] (li1) at (-0.75, 0) {};
                \node [style=none] (li1l) at (-0.9, 0) {A};
		\node [style=none] (ri1) at (0.75, 0.25) {};
		\node [style=none] (ri2) at (0.75, -0.25) {};
                \node [style=none] (ri1l) at (0.9, 0.25) {A};
		\node [style=none] (ri2l) at (0.9, -0.25) {B};
                \node [style=none] (ra1) at (0, 0.25) {};
                \node [style=none] (ra2) at (0, -0.25) {};                
	\end{pgfonlayer}
	\begin{pgfonlayer}{edgelayer}
		\draw (li1.center) to (r);
		\draw (ri1.center) to (ra1);
		\draw (ri2.center) to (ra2);
	\end{pgfonlayer}
\end{tikzpicture}

\begin{tikzpicture}
	\begin{pgfonlayer}{nodelayer}
		\node [style=smallrect] (r) at (0, 0) {)};
		\node [style=none] (li1) at (0.75, 0) {};
                \node [style=none] (li1l) at (0.9, 0) {A};
		\node [style=none] (ri1) at (-0.75, 0.25) {};
		\node [style=none] (ri2) at (-0.75, -0.25) {};
                \node [style=none] (ri1l) at (-0.9, 0.25) {A};
		\node [style=none] (ri2l) at (-0.9, -0.25) {B};
                \node [style=none] (ra1) at (0, 0.25) {};
                \node [style=none] (ra2) at (0, -0.25) {};                
	\end{pgfonlayer}
	\begin{pgfonlayer}{edgelayer}
		\draw (li1.center) to (r);
		\draw (ri1.center) to (ra1);
		\draw (ri2.center) to (ra2);
	\end{pgfonlayer}
\end{tikzpicture}

\begin{tikzpicture}
	\begin{pgfonlayer}{nodelayer}
		\node [style=none] (0) at (0.5, 0) {};
		\node [style=none] (1) at (0.2, 0) {};
                \node [style=none] (2) at (0, 0) {A};
	\end{pgfonlayer}
	\begin{pgfonlayer}{edgelayer}
		\draw [style=new edge style 0, in=180, out=0] (1.center) to (0.center);
	\end{pgfonlayer}
\end{tikzpicture}

   \hspace*{10mm} \begin{tikzpicture}
	\begin{pgfonlayer}{nodelayer}
		\node [style=smallrect] (r) at (0, 0) {(};
		\node [style=none] (li1) at (-0.5, 0) {};
                \node [style=none] (li1l) at (-0.9, 0) {};
		\node [style=none] (ri1) at (0.75, 0.25) {};
		\node [style=none] (ri2) at (0.75, -0.25) {};
                \node [style=none] (ri1l) at (0.9, 0.25) {};
		\node [style=none] (ri2l) at (0.9, -0.25) {};
                \node [style=none] (ra1) at (0, 0.25) {};
                \node [style=none] (ra2) at (0, -0.25) {};

                \node [style=smallrect] (Cr) at (1, 0.25) {(};
		\node [style=none] (Cli1) at (-0.75, 0.25) {};
                \node [style=none] (Cli1l) at (-0.9, 0.25) {};
		\node [style=none] (Cri1) at (0.75, 0.25) {};
		\node [style=none] (Cri2) at (0.75, -0.25) {};
                \node [style=none] (Cri1l) at (0.9, 0.25) {};
		\node [style=none] (Cri2l) at (0.9, -0.25) {};
                \node [style=none] (Cra1) at (0, 0.25) {};
                \node [style=none] (Cra2) at (0, -0.25) {};

                \node [style=smallrect] (Ar) at (2, 0.25) {)};
		\node [style=none] (Ali1) at (2.75, 0.25) {};
                \node [style=none] (Ali1l) at (2.9, 0) {};
		\node [style=none] (Ari1) at (1.25, 0.0) {};
                \node [style=none] (Arx1) at (1.25, 0.5) {};
		\node [style=none] (Ari2) at (1.25, -0.25) {};
                \node [style=none] (Ari1l) at (1.1, 0.25) {};
		\node [style=none] (Ari2l) at (1.1, -0.25) {};
                \node [style=none] (Ara1) at (2, 0.0) {};
                \node [style=none] (Arxa1) at (2, 0.5) {};
                \node [style=none] (Ara2) at (2, -0.25) {};         

                \node [style=smallrect] (Br) at (3, 0) {)};
		\node [style=none] (Bli1) at (3.5, 0) {};
                \node [style=none] (Bli1l) at (3.9, 0) {};
		\node [style=none] (Bri1) at (2.25, 0.25) {};
		\node [style=none] (Bri2) at (2.25, -0.25) {};
                \node [style=none] (Bri1l) at (2.1, 0.25) {};
		\node [style=none] (Bri2l) at (2.1, -0.25) {};
                \node [style=none] (Bra1) at (3, 0.25) {};
                \node [style=none] (Bra2) at (3, -0.25) {};
	\end{pgfonlayer}
	\begin{pgfonlayer}{edgelayer}
		\draw [style=new edge style 1, in=180, out=0] (li1) to (r.center);
                \draw (Ali1.center) to (Ar);

		\draw (Ari1.center) to (Ara1);
                \draw (Arx1.center) to (Arxa1);

		\draw (ra2.center) to (Bra2);

		\draw (Cri1.center) to (Cra1);
		
                \draw  [style=new edge style 0, in=180, out=0] (Br) to (Bli1.center);
		
	\end{pgfonlayer}
\end{tikzpicture}

  \end{center}
  This illustrates how regular monoidal grammars permit unbounded concurrency. Here, as one scans from left to right, the (unbounded) size of the internal boundary of a string diagram keeps track of the number of open left parentheses.
\end{example}

\begin{example}[Brick walls] \label{ex:brick}

  A variant on the ``brick wall'' language introduced by \cite{bossut} is given by the following grammar (left below). An example of a morphism in the language defined by this grammar is shown on the right.

  \begin{center}
    \begin{minipage}{.4\textwidth}\vspace*{-15mm}\begin{tikzpicture}
	\begin{pgfonlayer}{nodelayer}

                \node [style=none] (0) at (-1.8, 0.25) {};
                \node [style=none] (1) at (-1.5, 0.25) {};
                \node [style=none] (2) at (-1.3, 0.25) {$H$};
                \node [style=none] (3) at (-1.8, -0.25) {};
                \node [style=none] (4) at (-1.5, -0.25) {};
                \node [style=none] (5) at (-1.3, -0.25) {$V$};

		\node [style=smallrect] (r) at (0, 0) {};
		\node [style=none] (li1) at (-0.5, 0.25) {};
                \node [style=none] (li2) at (-0.5, -0.25) {};
                \node [style=none] (li1l) at (-0.7, 0.25) {$H$};
                \node [style=none] (li2l) at (-0.7, -0.25) {$V$};
                \node [style=none] (la1) at (0, 0.25) {};
                \node [style=none] (la2) at (0, -0.25) {};                
		\node [style=none] (ri1) at (0.5, 0.25) {};
		\node [style=none] (ri2) at (0.5, -0.25) {};
                \node [style=none] (ri1l) at (0.7, 0.25) {$V$};
		\node [style=none] (ri2l) at (0.7, -0.25) {$H$};
                \node [style=none] (ra1) at (0, 0.25) {};
                \node [style=none] (ra2) at (0, -0.25) {};

                \node [style=none] (6) at (1.7, 0.25) {};
                \node [style=none] (7) at (1.5, 0.25) {};
                \node [style=none] (8) at (1.3, 0.25) {$H$};
                \node [style=none] (9) at (1.7, -0.25) {};
                \node [style=none] (10) at (1.5, -0.25) {};
                \node [style=none] (11) at (1.3, -0.25) {$V$};
	\end{pgfonlayer}
	\begin{pgfonlayer}{edgelayer}
		\draw (li1.center) to (la1);
                \draw (li2.center) to (la2);
		\draw (ri1.center) to (ra1);
		\draw (ri2.center) to (ra2);

                \draw [style=new edge style 1] (0.center) to (1.center);
                \draw [style=new edge style 1] (3.center) to (4.center);
                \draw [style=new edge style 0] (7.center) to (6.center);
                \draw [style=new edge style 0] (10.center) to (9.center);
	\end{pgfonlayer}
\end{tikzpicture} \end{minipage} \resizebox{\height}{1.8cm}{\includegraphics[scale=0.7]{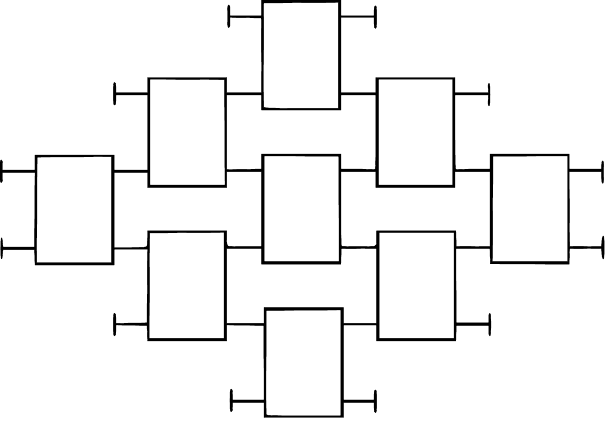}}
    \end{center}
\end{example}

In Section \ref{sec:closure} we will see how this language of ``brick walls'' allows us to construct the following example as an intersection of two languages:

\begin{example}[Sierpi\'nski gasket] \label{ex:sierpinski}
  In \cite{10.1371/journal.pbio.0020424}, self-assembly of DNA tiles was used to realize the behaviour of a cellular automaton that computes the Sierpi\'nski gasket fractal, based on the computation of the XOR gate. \cite{10.1371/journal.pbio.0020424} implicitly depicts a monoidal grammar, and so Sierpi\'nski gaskets of arbitrary iteration depth (e.g. right below) are in fact the monoidal language over this grammar (left below, where we use colours for the alphabet):
  \\ \\
  \begin{minipage}{.69\textwidth}\vspace*{-23mm}\begin{tikzpicture}
        \begin{pgfonlayer}{nodelayer}
                \node [style=none] (0) at (0.7, 0.25) {$H_1$};
                \node [style=none] (1) at (0.3, 0.25) {};
                \node [style=none] (2) at (0.5, 0.25) {};

                \node [style=none] (3) at (0.7, -0.25) {$V_1$};
                \node [style=none] (4) at (0.3, -0.25) {};
                \node [style=none] (5) at (0.5, -0.25) {};
        \end{pgfonlayer}
        \begin{pgfonlayer}{edgelayer}
                \draw [style=new edge style 0] (2.center) to (1.center);
                \draw [style=new edge style 0] (5.center) to (4.center);
        \end{pgfonlayer}
\end{tikzpicture} \begin{tikzpicture}
	\begin{pgfonlayer}{nodelayer}
		\node [style=smallrect, fill=cyan] (r) at (0, 0) {};
		\node [style=none] (li1) at (-0.5, 0.25) {};
                \node [style=none] (li2) at (-0.5, -0.25) {};
                \node [style=none] (li1l) at (-0.7, 0.25) {$H_1$};
                \node [style=none] (li2l) at (-0.7, -0.25) {$V_1$};
                \node [style=none] (la1) at (0, 0.25) {};
                \node [style=none] (la2) at (0, -0.25) {};                
		\node [style=none] (ri1) at (0.5, 0.25) {};
		\node [style=none] (ri2) at (0.5, -0.25) {};
                \node [style=none] (ri1l) at (0.7, 0.25) {$V_0$};
		\node [style=none] (ri2l) at (0.7, -0.25) {$H_0$};
                \node [style=none] (ra1) at (0, 0.25) {};
                \node [style=none] (ra2) at (0, -0.25) {};

                \node [style=smallrect, fill=cyan] (Ar) at (2, 0) {};
		\node [style=none] (Ali1) at (1.5, 0.25) {};
                \node [style=none] (Ali2) at (1.5, -0.25) {};
                \node [style=none] (Ali1l) at (1.3, 0.25) {$H_0$};
                \node [style=none] (Ali2l) at (1.3, -0.25) {$V_0$};
                \node [style=none] (Ala1) at (2, 0.25) {};
                \node [style=none] (Ala2) at (2, -0.25) {};
		\node [style=none] (Ari1) at (2.5, 0.25) {};
		\node [style=none] (Ari2) at (2.5, -0.25) {};
                \node [style=none] (Ari1l) at (2.7, 0.25) {$V_0$};
		\node [style=none] (Ari2l) at (2.7, -0.25) {$H_0$};
                \node [style=none] (Ara1) at (2, 0.25) {};
                \node [style=none] (Ara2) at (2, -0.25) {};

                \node [style=smallrect, fill=yellow] (Br) at (4, 0) {};
		\node [style=none] (Bli1) at (3.5, 0.25) {};
                \node [style=none] (Bli2) at (3.5, -0.25) {};
                \node [style=none] (Bli1l) at (3.3, 0.25) {$H_0$};
                \node [style=none] (Bli2l) at (3.3, -0.25) {$V_1$};
                \node [style=none] (Bla1) at (4, 0.25) {};
                \node [style=none] (Bla2) at (4, -0.25) {};
		\node [style=none] (Bri1) at (4.5, 0.25) {};
		\node [style=none] (Bri2) at (4.5, -0.25) {};
                \node [style=none] (Bri1l) at (4.7, 0.25) {$V_1$};
		\node [style=none] (Bri2l) at (4.7, -0.25) {$H_1$};
                \node [style=none] (Bra1) at (4, 0.25) {};
                \node [style=none] (Bra2) at (4, -0.25) {};

                \node [style=smallrect, fill=yellow] (Cr) at (6, 0) {};
		\node [style=none] (Cli1) at (5.5, 0.25) {};
                \node [style=none] (Cli2) at (5.5, -0.25) {};
                \node [style=none] (Cli1l) at (5.3, 0.25) {$H_1$};
                \node [style=none] (Cli2l) at (5.3, -0.25) {$V_0$};
                \node [style=none] (Cla1) at (6, 0.25) {};
                \node [style=none] (Cla2) at (6, -0.25) {};
		\node [style=none] (Cri1) at (6.5, 0.25) {};
		\node [style=none] (Cri2) at (6.5, -0.25) {};
                \node [style=none] (Cri1l) at (6.7, 0.25) {$V_1$};
		\node [style=none] (Cri2l) at (6.7, -0.25) {$H_1$};
                \node [style=none] (Cra1) at (6, 0.25) {};
                \node [style=none] (Cra2) at (6, -0.25) {};
	\end{pgfonlayer}
	\begin{pgfonlayer}{edgelayer}
		\draw (li1.center) to (la1);
                \draw (li2.center) to (la2);
		\draw (ri1.center) to (ra1);
		\draw (ri2.center) to (ra2);
                
                \draw (Ali1.center) to (Ala1);
                \draw (Ali2.center) to (Ala2);
		\draw (Ari1.center) to (Ara1);
		\draw (Ari2.center) to (Ara2);

                \draw (Bli1.center) to (Bla1);
                \draw (Bli2.center) to (Bla2);
		\draw (Bri1.center) to (Bra1);
		\draw (Bri2.center) to (Bra2);

                \draw (Cli1.center) to (Cla1);
                \draw (Cli2.center) to (Cla2);
		\draw (Cri1.center) to (Cra1);
		\draw (Cri2.center) to (Cra2);
	\end{pgfonlayer}
\end{tikzpicture} \begin{tikzpicture}
        \begin{pgfonlayer}{nodelayer}
                \node [style=none] (0) at (0, 0.25) {$V_1$};
                \node [style=none] (1) at (0.2, 0.25) {};
                \node [style=none] (2) at (0.4, 0.25) {};

                \node [style=none] (3) at (0, -0.25) {$H_1$};
                \node [style=none] (4) at (0.2, -0.25) {};
                \node [style=none] (5) at (0.4, -0.25) {};
        \end{pgfonlayer}
        \begin{pgfonlayer}{edgelayer}
                \draw [style=new edge style 0] (4.center) to (5.center);
                \draw [style=new edge style 0] (1.center) to (2.center);
        \end{pgfonlayer}
\end{tikzpicture}\end{minipage} ~ \includegraphics[width=0.295\textwidth]{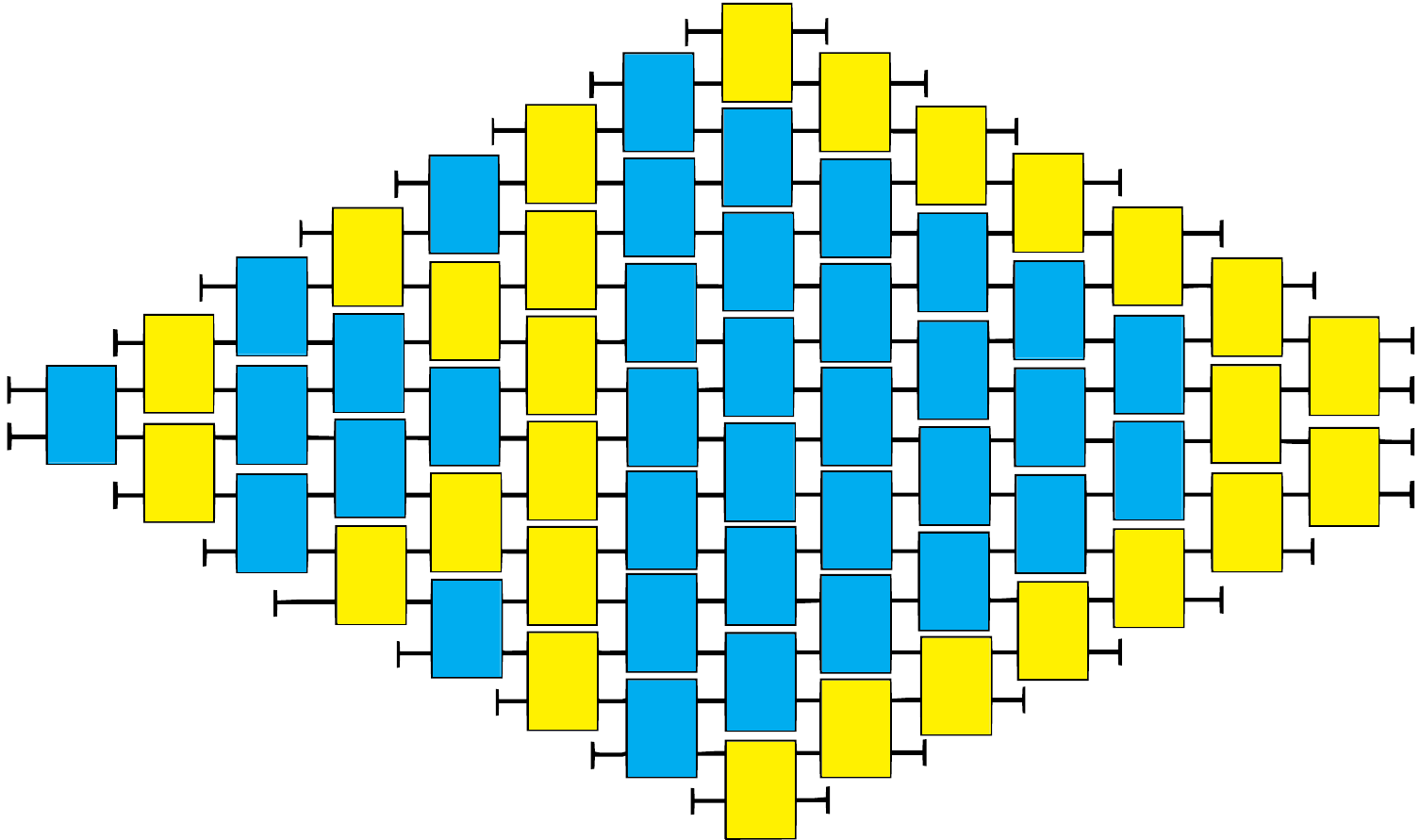}
\end{example}

\begin{example} \label{ex:nondet}
  We define a grammar (left below) that will serve as a running counterexample in Section \ref{sec:determinizability}, as it defines a language that cannot be deterministically recognized. The connected string diagrams in this language are exactly two (right below).

  \begin{center}
    \includegraphics[scale=0.55]{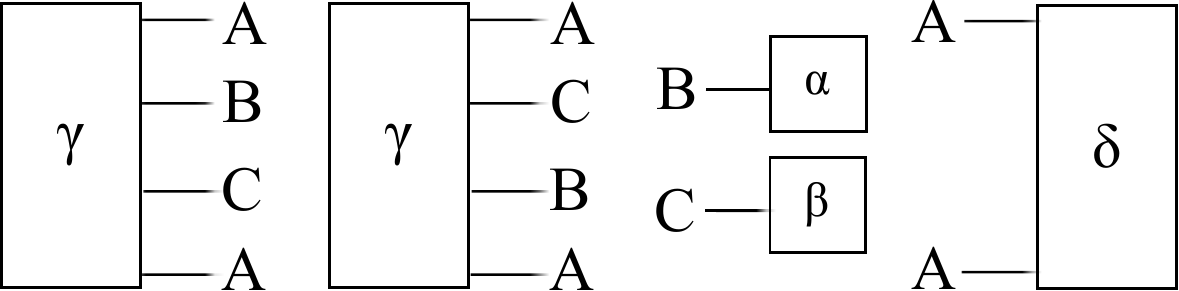} \hspace*{10mm}  \includegraphics[scale=0.55]{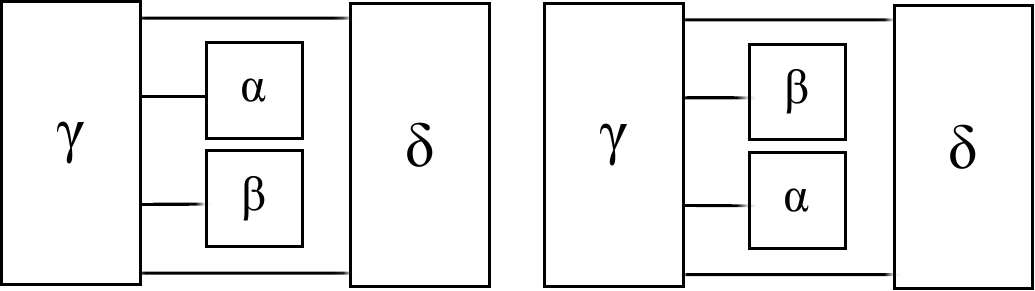}
  \end{center}
\end{example}

\begin{remark}
If the monoidal graph $\mathcal{M}$ has no edges whose domain is $\varepsilon$ and no edges whose codomain is $\varepsilon$, a regular monoidal grammar $\Psi : \mathcal{M} \to \Gamma$ will define a language containing only the identity on the monoidal unit, i.e.\ the empty string diagram (denoted \raisebox{.25em}{\begin{tikzpicture}[scale=.5,baseline=(current bounding box.center)]
\akasa
\end{tikzpicture}}\!). In fact, \emph{every} monoidal language contains the empty string diagram.
\end{remark}

\subsection{Non-deterministic monoidal automata} \label{sec:pro_aut}

Recall that a non-deterministic finite automaton (NFA) is given by a finite set $Q$ of states, an initial state $i \in Q$, a set of final states $F \subseteq Q$, and for each $a \in \Sigma$, a function $Q \xrightarrow[]{\Delta_a} \mathscr{P}(Q)$. Non-deterministic \emph{monoidal automata} do not have initial and final states; string diagrams are simply accepted or rejected depending on their shape. In Section \ref{sec:classical}, we will see that initial and final states derive from this definition, when the alphabet is of a particular form.

\begin{definition} \label{defn:nd_aut}
A non-deterministic monoidal automaton $\Delta = (Q, \Delta_\Gamma)$ over a monoidal alphabet $\Gamma$ is given by a finite set $Q$, together with a set of transition functions indexed by generators $\Delta_\Gamma = \{Q^{\text{ar}(\gamma)} \xrightarrow[]{\Delta_\gamma} \mathscr{P}(Q^{\text{coar}(\gamma)})\}_{\gamma \in E_\Gamma}$.
\end{definition}

For classical NFAs, the assignment $a \mapsto \Delta_a$ extends uniquely to a functor $\Sigma^{*} \to \textsf{Rel}$, the inductive extension of the transition structure from letters to words. We define the inductive extension of monoidal automata from generators to string diagrams. First recall the definition of the endomorphism pro of an object in a monoidal category:

\begin{definition}
  Let $\mathscr{C}$ be a monoidal category, and $Q$ an object of $\mathscr{C}$. The endomorphism pro of $Q$, $\endo{Q}$, has natural numbers as objects, hom-sets $\endo{Q}(n,m) := \mathscr{C}(Q^n, Q^m)$, composition and identities as in $\mathscr{C}$. The monoidal product is addition on objects, and as in $\mathscr{C}$ on morphisms.
\end{definition}

The codomains of our inductive extension will be endomorphism pros of finite sets $Q$ in $\Rel$, considered as the Kleisli category of the powerset monad $\mathscr{P}$. Since $\mathscr{P}$ is a commutative monad (with respect to the cartesian product of sets, with $\mathscr{P}X \times \mathscr{P}Y \to \mathscr{P}(X \times Y)$ given by the product of subsets), the following lemma gives us the monoidal structure on $\Rel$:

\begin{lemma}[\cite{power_robinson_1997}, Corollary 4.3] \label{lemma:comm_monad}
  Let $T$ be a commutative monad on a symmetric monoidal category $\mathscr{C}$. Then the Kleisli category $\textsf{Kl}(T)$ has a canonical monoidal structure, which is given on objects by the monoidal product in $\mathscr{C}$, and on morphisms $f : X \to TA, g : Y \to TB$ by $X \otimes Y \xrightarrow[]{f \otimes g} TA \otimes TB \xrightarrow[]{\nabla} T(A \otimes B)$, where $\nabla$ is 
  given by the commutativity of $T$.
\end{lemma}

\begin{remark} \label{remark:par}
  The maybe monad $(\text{--})_\bot$ is also commutative, so its Kleisli category, equivalent to the category $\Par$ of sets and partial functions, also has a canonical monoidal structure, and for each set $Q$ there is an endomorphism pro $\Par[Q]$. We will come back to $\Par[Q]$ in Section \ref{sec:deterministic}.
\end{remark}

Now we can define the inductive extension of a non-deterministic monoidal automaton:

\begin{observation} \label{obs:free}
  The assignment of generators to transition functions $\gamma \mapsto \Delta_\gamma$ in Definition \ref{defn:nd_aut} determines a morphism of monoidal graphs $\Gamma \to |\Rel_Q|$. Such morphisms are in bijection with pro morphisms $\Delta : \mathscr{F}\Gamma \to \Rel[Q]$. We will also refer to the inductive extension $\Delta$ as a non-deterministic monoidal automaton, and sometimes write $\Delta_\alpha$ for the relation $\Delta(\alpha : n \to m)$.
\end{observation}

A scalar string diagram is mapped to one of the two possible nullary relations $\{\bullet\} \to \mathscr{P}(\{\bullet\})$, which represent accepting or rejecting computations, and thus can be used to define the language of the automaton:

\begin{definition} \label{defn:lang_ndet}
  Let $\Delta : \mathscr{F}\Gamma \to \Rel[Q]$ be a non-deterministic monoidal automaton. Then the monoidal language accepted by $\Delta$ is $\Lang(\Delta) := \{ \alpha \in \mathscr{F}\Gamma(0, 0) \mid \Delta_\alpha(\bullet) = \{\bullet\} \}$.
\end{definition}

There is an evident correspondence between regular monoidal grammars and non-deterministic monoidal automata. The graphical representation of a grammar makes this most clear: it can also be thought of as the ``transition graph'' of a non-deterministic monoidal automaton. More explicitly we have:

\begin{proposition}
  Given a regular monoidal grammar $\Psi : \mathcal{M} \to \Gamma$, define a monoidal automaton with $Q = V_{\mathcal{M}}$, $w(\Delta_\gamma)w' \iff \exists \sigma \in E_\Psi^{-1}(\gamma)$ such that $dom(\sigma) = w, cod(\sigma) = w'$. Conversely given a monoidal automaton $(Q,\Delta_\Gamma)$, define a regular monoidal grammar with $V_{\mathcal{M}} = Q$ and take an edge $w \to w'$ over $\gamma$ $\iff$ $w(\Delta_\gamma)w'$. This correspondence of grammars and automata preserves the recognized language.
\end{proposition}

\begin{remark}
  In automata theory it is often convenient to consider automata with $\varepsilon$-transitions, or word-labelled transitions more generally. As monoidal grammars, these correspond to arbitrary functors $\mathscr{F}\mathcal{M} \to \mathscr{F}\Gamma$, that is (by the adjunction $\mathscr{U} \dashv \mathscr{F}$), to morphisms of finite monoidal graphs $\mathcal{M} \to \mathscr{UF}\Gamma$. The corresponding generalization of monoidal automata requires considering $\Rel[Q]$ as a monoidal 2-category with 2-cells the inclusions. Identity on objects, strict monoidal lax 2-functors $\mathscr{F}\Gamma \to \Rel[Q]$ (where $\mathscr{F}\Gamma$ is considered as equipped with identity 2-cells), then give the refined notion of monoidal automaton. Such a lax 2-functor need no longer send the identity on $n$ wires to the identity relation on $Q^n$, but merely to a relation that includes the identity; this corresponds to allowing silent transitions. Similarly, \emph{lax} preservation of composition corresponds to allowing ``diagram-labelled'' transitions.
\end{remark}

\subsection{Closure properties of regular monoidal languages} \label{sec:closure}
We record some closure properties of regular monoidal languages.

\begin{lemma}[Closure under union] \label{lemma:closed_union}
  Let $L$ and $L'$ be regular monoidal languages over $\Gamma$. Then $L \cup L'$ is a regular monoidal language over $\Gamma$.
\end{lemma}
\begin{proof}
  Let $L$ and $L'$ be given by the regular monoidal grammars $\Psi : \mathcal{M} \to \Gamma, \Psi' : \mathcal{M}' \to \Gamma$ respectively. Define the grammar $\Psi + \Psi' : \mathcal{M} + \mathcal{M}' \to \Gamma$, where $E_{\mathcal{M}+\mathcal{M}'} := E_{\mathcal{M}} + E_{\mathcal{M}'}, V_{\mathcal{M}+\mathcal{M'}} := V_{\mathcal{M}}+V_{\mathcal{M}'}$, and $E_{\Psi+\Psi'} := [\Psi, \Psi']$ (the copairing of $\Psi$ and $\Psi'$). Graphically, this is just taking the disjoint union of two grammars, and it is clear that the language defined in this way is the union of the languages defined by the two grammars.
\end{proof}

\begin{lemma}[Closure under intersection] \label{lemma:closed_intersection}
  Let $L$ and $L'$ be regular monoidal languages over $\Gamma$. Then $L \cap L'$ is a regular monoidal language over $\Gamma$.
\end{lemma}
\begin{proof}
  Let $L$ and $L'$ be recognized by non-deterministic automata $\Delta : \mathscr{F}\Gamma \to \Rel[Q], \Delta' : \mathscr{F}\Gamma \to \Rel[Q']$ respectively. Consider the product automaton $\Delta \times \Delta' : \mathscr{F}\Gamma \to \Rel[Q \times Q']$ defined by $(\Delta \times \Delta')_\alpha := (\Delta_\alpha \times \Delta'_\alpha) \comp \nabla$, where $\nabla$ is the monoidal multiplication given by the commutativity of the powerset monad. Then $\alpha$ is accepted by $\Delta \times \Delta'$ just when it is accepted by both, so $\Lang(\Delta \times \Delta') = L \cap L'$.
\end{proof}

\begin{remark}
  The Sierpi\'nski gasket language (Example \ref{ex:sierpinski}) is the intersection of the brick wall language (Example \ref{ex:brick}) and an ``XOR gate'' language: this explains the origin of the states in the grammar shown in Example \ref{ex:sierpinski}.
\end{remark}

\begin{lemma}[Closure under monoidal product and factors] \label{lemma:closed_monoidal_product}
Let $L$ be a regular monoidal language. Then $\alpha,\beta \in L \iff \alpha \otimes \beta \in L$.
\end{lemma}
\begin{proof}
  Let $\Delta : \mathscr{F}\Gamma \to \Rel[Q]$ be an automaton accepting both $\alpha$ and $\beta$. Since $\Delta$ is a strict monoidal functor, $\Delta(\alpha \otimes \beta) = \Delta(\alpha) \otimes \Delta(\beta)$, so we must have $\Delta(\alpha \otimes \beta)(\bullet) = \{\bullet\}$, and conversely.
\end{proof}

\begin{lemma}[Closure under images of alphabets] \label{lemma:closed_images}
  Let $L$ a be regular monoidal language over $\Gamma$, and $\Gamma \xrightarrow[]{h} \Gamma'$ be a morphism of monoidal alphabets. Then $(\mathscr{F}h)L$ is a regular monoidal language over $\Gamma'$.
\end{lemma}
\begin{proof}
Let $L$ be given by the regular monoidal grammar $\Psi : \mathcal{M} \to \Gamma$, that is $L = \mathscr{F}\Psi[\mathscr{F}\mathcal{M}(\varepsilon, \varepsilon)]$. Consider the grammar given by the composite $\Psi \comp h : \mathcal{M} \to \Gamma'$. Since $\mathscr{F}$ is a functor we have: $\mathscr{F}(\Psi \comp h)[\mathscr{F}\mathcal{M}(\varepsilon, \varepsilon)] = (\mathscr{F}\Psi \comp \mathscr{F}h)[\mathscr{F}\mathcal{M}(\varepsilon, \varepsilon)] = (\mathscr{F}h)L$, thus $\Psi \comp h$ is a grammar for $(\mathscr{F}h)L$.
\end{proof}

\begin{lemma}[Closure under preimages of alphabets] \label{lemma:closed_preimages}
  Let $L$ a regular monoidal language over $\Gamma$, and $\Gamma' \xrightarrow[]{h} \Gamma$ be a morphism of monoidal alphabets. Then the inverse image of L, $(\mathscr{F}h)^{-1}(L)$ is a regular monoidal language over $\Gamma'$.
\end{lemma}
\begin{proof}
  Let $\Delta : \mathscr{F}\Gamma \to \Rel[Q]$ be an automaton recognizing $L$. Consider the automaton given by the composite $\mathscr{F}h \comp \Delta : \mathscr{F}\Gamma' \to \Rel[Q]$. We have $\Lang(\mathscr{F}h \comp \Delta) = (\mathscr{F}h)^{-1}(\Lang(\Delta)) = (\mathscr{F}h)^{-1}(L)$, so the inverse image of $L$ is regular.
\end{proof}

Closure under complement is often held to be an important criterion for what should count as a \emph{recognizable} language. Indeed, for the abstract monadic second order logic introduced in \cite{mmso}, it is a \emph{theorem} that the class of recognizable languages relative to a monad on \textsf{Set} is closed under complement.
However, given that every monoidal language contains the empty string diagram, we obviously have that:

\begin{observation} \label{lemma:nonclosed_complement}
  Regular monoidal languages are not closed under complement.
\end{observation}

This suggests that there is no obvious account of regular monoidal languages in terms of monadic second order logic. On the other hand, there is no reason we should expect even the general account of monadic second order logic given in \cite{mmso} to extend to monoidal categories, since these are not algebras for a monad on \textsf{Set}. Moreover, taking inspiration from classical examples in Section~\ref{sec:classical}, one could also refine what is meant by complement, for instance focussing on the set of non-empty connected scalar diagrams -- see below for more details.

\section{Regular word and tree languages as regular monoidal languages} \label{sec:classical}

Classical non-deterministic finite-state automata and tree automata can be seen as non-deterministic monoidal automata over alphabets of a particular shape.

To make the correspondence precise, in the following we restrict monoidal languages to their \emph{connected} string diagrams. Strictly speaking, the language of a monoidal automaton always contains only the empty diagram or is countably infinite, because if $\alpha$ is accepted by the automaton, so are arbitrary finite monoidal products $\alpha \otimes \dots \otimes \alpha$. However, it is of course possible for a monoidal language to consist of a finite number of connected string diagrams.

From another perspective, without restricting to connected components, we can say that the monoidal automata corresponding to finite-state and tree automata have the power of an unbounded number of such classical automata running in parallel.

\subsection{Finite-state automata} \label{example:reg_grammar}

\begin{definition} \label{defn:word_alphabet}
A word monoidal alphabet is a monoidal alphabet having only generators of arity and coarity 1, \providecommand{\generatorName}{}\renewcommand{\generatorName}{\sigma}\raisebox{0.1em}{\begin{tikzpicture}[every node/.style={inner sep=0,outer sep=0}]
	\begin{pgfonlayer}{nodelayer}
		\node [style=new style 0] (2) at (0, 0) {$\generatorName$};
		\node [style=none] (3) at (0.5, 0) {};
		\node [style=none] (4) at (-0.5, 0) {};
	\end{pgfonlayer}
	\begin{pgfonlayer}{edgelayer}
		\draw (3.center) to (2);
		\draw (4.center) to (2);
	\end{pgfonlayer}
\end{tikzpicture}
}{}, along with a single ``start'' generator ~\raisebox{0.1em}{\begin{tikzpicture}
	\begin{pgfonlayer}{nodelayer}
		\node [style=none] (2) at (0, 0) {};
		\node [style=none] (3) at (0.4, 0) {};
	\end{pgfonlayer}
	\begin{pgfonlayer}{edgelayer}
		\draw [style=new edge style 1] (2.center) to (3);
	\end{pgfonlayer}
\end{tikzpicture}
} of arity 0 and coarity 1, and ``end'' generator \raisebox{0.1em}{\begin{tikzpicture}
	\begin{pgfonlayer}{nodelayer}
		\node [style=none] (2) at (0, 0) {};
		\node [style=none] (3) at (0.4, 0) {};
	\end{pgfonlayer}
	\begin{pgfonlayer}{edgelayer}
		\draw [style=new edge style 0] (2.center) to (3);
	\end{pgfonlayer}
\end{tikzpicture}
}\! of arity 1 and coarity 0.
\end{definition}

\begin{observation}
 Non-deterministic monoidal automata over word monoidal alphabets correspond to classical NFAs.
\end{observation}
  Let an NFA $A = (Q, \Sigma, \Delta, i, F)$ be given. We build a monoidal automaton as follows. Form the monoidal alphabet $\Sigma'$ by starting with generators \raisebox{0.1em}{\begin{tikzpicture}
	\begin{pgfonlayer}{nodelayer}
		\node [style=none] (2) at (0, 0) {};
		\node [style=none] (3) at (0.4, 0) {};
	\end{pgfonlayer}
	\begin{pgfonlayer}{edgelayer}
		\draw [style=new edge style 1] (2.center) to (3);
	\end{pgfonlayer}
\end{tikzpicture}
}, \raisebox{0.1em}{\begin{tikzpicture}
	\begin{pgfonlayer}{nodelayer}
		\node [style=none] (2) at (0, 0) {};
		\node [style=none] (3) at (0.4, 0) {};
	\end{pgfonlayer}
	\begin{pgfonlayer}{edgelayer}
		\draw [style=new edge style 0] (2.center) to (3);
	\end{pgfonlayer}
\end{tikzpicture}
}\!
  and adding generators \providecommand{\generatorName}{}\renewcommand{\generatorName}{\sigma}\raisebox{0.1em}{\begin{tikzpicture}[every node/.style={inner sep=0,outer sep=0}]
	\begin{pgfonlayer}{nodelayer}
		\node [style=new style 0] (2) at (0, 0) {$\generatorName$};
		\node [style=none] (3) at (0.5, 0) {};
		\node [style=none] (4) at (-0.5, 0) {};
	\end{pgfonlayer}
	\begin{pgfonlayer}{edgelayer}
		\draw (3.center) to (2);
		\draw (4.center) to (2);
	\end{pgfonlayer}
\end{tikzpicture}
} for each $\sigma \in \Sigma$.  For each \providecommand{\generatorName}{}\renewcommand{\generatorName}{\sigma}\raisebox{0.1em}{\begin{tikzpicture}[every node/.style={inner sep=0,outer sep=0}]
	\begin{pgfonlayer}{nodelayer}
		\node [style=new style 0] (2) at (0, 0) {$\generatorName$};
		\node [style=none] (3) at (0.5, 0) {};
		\node [style=none] (4) at (-0.5, 0) {};
	\end{pgfonlayer}
	\begin{pgfonlayer}{edgelayer}
		\draw (3.center) to (2);
		\draw (4.center) to (2);
	\end{pgfonlayer}
\end{tikzpicture}
}, take the transition function $\Delta_\sigma := \Delta(\sigma, \text{--}) : Q \to \mathscr{P}(Q)$. For \raisebox{0.1em}{\begin{tikzpicture}
	\begin{pgfonlayer}{nodelayer}
		\node [style=none] (2) at (0, 0) {};
		\node [style=none] (3) at (0.4, 0) {};
	\end{pgfonlayer}
	\begin{pgfonlayer}{edgelayer}
		\draw [style=new edge style 0] (2.center) to (3);
	\end{pgfonlayer}
\end{tikzpicture}
} take the transition function $Q \to \mathscr{P}(Q^0)$ to be the characteristic function of $F \subseteq Q$, sending elements of $F$ to $\{ \bullet \}$ and to $\varnothing$ otherwise, and for \raisebox{0.1em}{\begin{tikzpicture}
	\begin{pgfonlayer}{nodelayer}
		\node [style=none] (2) at (0, 0) {};
		\node [style=none] (3) at (0.4, 0) {};
	\end{pgfonlayer}
	\begin{pgfonlayer}{edgelayer}
		\draw [style=new edge style 1] (2.center) to (3);
	\end{pgfonlayer}
\end{tikzpicture}
} take the function $Q^0\to \mathscr{P}(Q)$ to pick out the singleton $\{i\}$. This defines a monoidal automaton $A' := (Q, {\Delta'}_{\Sigma'})$, and a simple induction shows that $\Lang(A) = \Lang(A')$, if one restricts to connected string diagrams.

Conversely, the data of a monoidal automaton over a word monoidal alphabet corresponds to the data of an NFA, the only difference being that the transition function associated to \raisebox{0.1em}{\begin{tikzpicture}
	\begin{pgfonlayer}{nodelayer}
		\node [style=none] (2) at (0, 0) {};
		\node [style=none] (3) at (0.4, 0) {};
	\end{pgfonlayer}
	\begin{pgfonlayer}{edgelayer}
		\draw [style=new edge style 1] (2.center) to (3);
	\end{pgfonlayer}
\end{tikzpicture}
} picks out a \emph{set} of initial states $\{\bullet\} \to \mathscr{P}(Q)$. We can always ``normalize'' such an automaton into an equivalent NFA with one initial state (see \cite[\S 2.3.1]{sakarovitch2009elements}). This shows how NFA initial and final states are captured by this particular shape of monoidal alphabet.

\subsection{Tree automata} \label{sec:tree_grammar}
Recall that non-deterministic finite tree automata come in two flavours, bottom-up and top-down, depending on whether they process a tree starting at the leaves or at the root, respectively. A non-deterministic bottom-up finite tree automaton is given by a finite set of states $Q$, a ``ranked'' alphabet $(\Sigma, r : \Sigma \to \mathbb{N})$, a set of final states $F \subseteq Q$, and for each $\sigma \in \Sigma$ a transition function $\Delta_\sigma : Q^{r(\sigma)} \to \mathscr{P}(Q)$. A non-deterministic top-down tree automaton, instead, has a set of initial states $I \subseteq Q$ and transition functions $\Delta_\sigma : Q \to \mathscr{P}(Q^{r(\sigma)})$. We can recover these as non-deterministic monoidal automata over \emph{tree monoidal alphabets}:

\begin{definition} \label{defn:tree_alphabet}
  A top-down tree monoidal alphabet is a monoidal alphabet having only generators of arity 1 (and arbitrary coarities $\geqslant 0$), \providecommand{\generatorName}{}\renewcommand{\generatorName}{\sigma}\raisebox{0em}{\begin{tikzpicture}[every node/.style={inner sep=0,outer sep=0}]
	\begin{pgfonlayer}{nodelayer}
		\node [style=new style 0] (2) at (0, 0) {$\generatorName$};
                \node [style=none,scale=0.75] at (0.5,0.08) {\Large $\vdots$};
                \node [style=none] (5) at (0.5, 0.2) {};
                \node [style=none] (7) at (0.5, -0.2) {};
		\node [style=none] (4) at (-0.5, 0) {};
	\end{pgfonlayer}
	\begin{pgfonlayer}{edgelayer}
                \draw (5.center) to (2);
                \draw (7.center) to (2);
		\draw (4.center) to (2);
	\end{pgfonlayer}
\end{tikzpicture}}, along with a single ``root'' generator ~\raisebox{0.1em}{\begin{tikzpicture}
	\begin{pgfonlayer}{nodelayer}
		\node [style=none] (2) at (0, 0) {};
		\node [style=none] (3) at (0.4, 0) {};
	\end{pgfonlayer}
	\begin{pgfonlayer}{edgelayer}
		\draw [style=new edge style 1] (2.center) to (3);
	\end{pgfonlayer}
\end{tikzpicture}
}\!. Analogously, a bottom-up tree monoidal alphabet is a monoidal alphabet having only generators of coarity 1 (and arbitrary arities $\geqslant 0$), \providecommand{\generatorName}{}\renewcommand{\generatorName}{\sigma}\raisebox{0em}{\begin{tikzpicture}[every node/.style={inner sep=0,outer sep=0}]
	\begin{pgfonlayer}{nodelayer}
		\node [style=new style 0] (2) at (0, 0) {$\generatorName$};
                \node [style=none,scale=0.75] at (-0.5,0.08) {\Large $\vdots$};
                \node [style=none] (5) at (-0.5, 0.2) {};
                \node [style=none] (7) at (-0.5, -0.2) {};
		\node [style=none] (4) at (0.5, 0) {};
	\end{pgfonlayer}
	\begin{pgfonlayer}{edgelayer}
                \draw (5.center) to (2);
                \draw (7.center) to (2);
		\draw (4.center) to (2);
	\end{pgfonlayer}
\end{tikzpicture}}, along with a single ``root'' generator \raisebox{0.1em}{\begin{tikzpicture}
	\begin{pgfonlayer}{nodelayer}
		\node [style=none] (2) at (0, 0) {};
		\node [style=none] (3) at (0.4, 0) {};
	\end{pgfonlayer}
	\begin{pgfonlayer}{edgelayer}
		\draw [style=new edge style 0] (2.center) to (3);
	\end{pgfonlayer}
\end{tikzpicture}
}\!.
\end{definition}

\begin{observation}  \label{ex:tree_aut}
Bottom-up tree automata are exactly non-deterministic monoidal automata over bottom-up tree monoidal alphabets, and likewise for top-down tree automata.
\end{observation}

The idea is similar to that sketched above for NFAs. For example, consider the following graph of a monoidal automaton over a bottom-up tree monoidal alphabet, recognizing trees corresponding to terms of the inductive type of lists of boolean values (a list may be empty, $[]$, or be a boolean value ``consed'' onto a list via $::$).

\begin{center}
\includegraphics[scale=0.5]{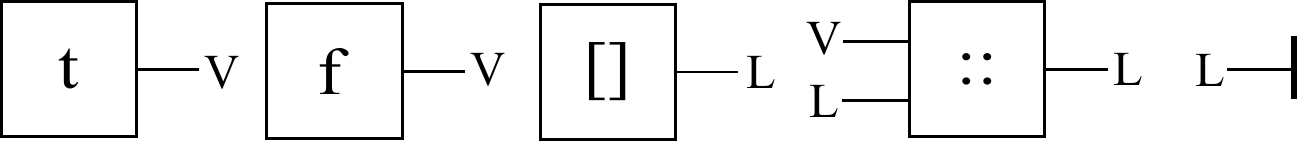}
\end{center}

Intuitively, the connected scalar string diagrams determined by this language are trees, with leaves on the left, and the root on the right. Monoidal automata over top-down tree monoidal alphabets have a similar form, but are mirrored horizontally, and thus morphisms in the language have the root on the left, and leaves on the right, and monoidal automata read the morphism starting at the root.

\section{The syntactic pro of a monoidal language} \label{sec:algebraic}
In this section we introduce the \emph{syntactic congruence} on monoidal languages and the corresponding \emph{syntactic pro}, by analogy with the syntactic congruence on classical regular languages and their associated syntactic monoid. In Section \ref{sec:sufficient} we will give an algebraic property of the syntactic pro sufficient for the language to be deterministically recognizable.

\begin{definition} \label{defn:context}
  A context of capacity $(n,m)$, where $n,m \geqslant 0$, is a scalar string diagram with a hole -- as illustrated below -- with zero or more additional wires exiting the first box and entering the second (indicated by ellipses).
  \begin{center}
    \includegraphics{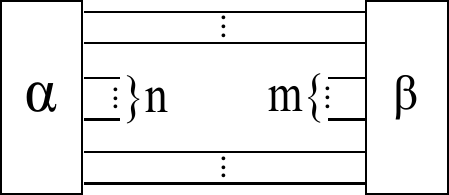}
  \end{center}
\end{definition}

 Given a context of capacity $(n,m)$, we can fill the hole with a string diagram $\alpha : n \to m$. Write $C[\alpha]$ for the resulting string diagram.

 Note that the empty diagram is a context, the empty context. Contexts allow us to define contextual equivalence of string diagrams:

\begin{definition}[Syntactic congruence] \label{defn:syntactic_cong}
  Given a monoidal language $L \subseteq \mathscr{F}\Gamma(0,0)$ we define its syntactic congruence $\equiv_L$ as follows. Let $\alpha, \beta$ be morphisms in $\mathscr{F}\Gamma(n,m)$. Then $\alpha \equiv_L \beta$ whenever $C[\alpha] \in L \iff C[\beta] \in L$, for all contexts $C$ of capacity $(n,m)$.
\end{definition}

\begin{definition} \label{defn:syntactic_pro}
  The syntactic pro of a monoidal language $L$ is the quotient pro $\mathscr{F}\Gamma /{\equiv_L}$. The quotient functor $S_L : \mathscr{F}\Gamma \to \mathscr{F}\Gamma/{\equiv_L}$ is the syntactic morphism of $L$. See Appendix \ref{appendix:congruence} for the definition of quotient pro and quotient functor.
\end{definition}

\begin{remark}
  The syntactic congruences for classical regular languages of words and trees are also special cases of this congruence over word and tree monoidal alphabets.
\end{remark}

\begin{lemma} \label{lemma:recog_via_empty}
$L$ is the inverse image along the syntactic morphism of the equivalence class of the empty diagram.
\end{lemma}
\begin{proof}
Let $\alpha \in L$. Then $\alpha \equiv_L \raisebox{.25em}{\begin{tikzpicture}[scale=.5,baseline=(current bounding box.center)]
\akasa
\end{tikzpicture}}$, since the empty diagram is in every language and if $C$ is a context of capacity $(0,0)$ distinguishing $\alpha$ and $\raisebox{.25em}{\begin{tikzpicture}[scale=.5,baseline=(current bounding box.center)]
\akasa
\end{tikzpicture}}$, then we have a contradiction by Lemma \ref{lemma:closed_monoidal_product}. So $\alpha \in S_L^{-1}(\left[\raisebox{.25em}{\begin{tikzpicture}[scale=.5,baseline=(current bounding box.center)]
\akasa
\end{tikzpicture}}\!\!\right])$, and conversely.
\end{proof}

In the terminology of algebraic language theory, we say that the syntactic morphism \emph{recognizes} $L$. A full investigation of algebraic recognizability of monoidal languages is a topic for future work. For now, we record the following lemma which is needed for Theorem \ref{thm:sufficient}:

\begin{lemma} \label{lemma:reg_then_lf}
  If a monoidal language L is regular, then its syntactic pro $\mathscr{F}\Gamma / {\equiv_L}$ is locally finite (i.e. has finite hom-sets).
\end{lemma}
\begin{proof}
  It suffices to exhibit a full pro morphism into $\mathscr{F}\Gamma/{\equiv_L}$ from a locally finite pro. Let $L$ be a regular monoidal language recognized by $\Delta : \mathscr{F}\Gamma \to \Rel[Q]$. $\Delta$ induces a congruence $\sim$ on $\mathscr{F}\Gamma$ defined by $\alpha \sim \beta \iff \Delta(\alpha) = \Delta(\beta)$, which implies that $\mathscr{F}\Gamma/{\sim}$ is locally finite, since $\textsf{Rel}_Q$ is locally finite. Define the pro morphism $\mathscr{F}\Gamma/{\sim} \to \mathscr{F}\Gamma/{\equiv_L}$ to be identity on objects and $[\alpha]_{\sim} \mapsto [\alpha]_{\equiv_L}$ on morphisms. This is well-defined since if $\alpha \sim \beta$ and $C[\alpha] \in L$ for some context $C$, then by functoriality $C[\beta] \in L$. Clearly it is full, so $\mathscr{F}\Gamma / {\equiv_L}$ is locally finite.
\end{proof}

\section{Deterministic monoidal automata} \label{sec:deterministic}

Classically, the expressive equivalence of deterministic and non-deterministic finite-state automata for string languages is well known, but already for trees, top-down deterministic tree automata are less expressive than bottom-up deterministic tree automata. Therefore we cannot expect to determinize non-deterministic monoidal automata. However, we have already seen monoidal languages that are deterministically recognizable (Examples \ref{ex:brackets}, \ref{ex:brick}, \ref{ex:sierpinski}, interpreted as the transition relations of monoidal automata, are functional relations). Here we  introduce deterministic monoidal automata and show that their languages enjoy the property of \emph{causal closure}. In Section \ref{sec:determinizability} we consider the question of determinizability.

\begin{definition} \label{defn:det_aut}
A deterministic monoidal automaton $\delta = (Q, \delta_\Gamma)$ over a monoidal alphabet $\Gamma$ is given by a finite set $Q$, together with transition functions $\delta_\Gamma = \{ Q^{\text{ar}(\gamma)} \xrightarrow[]{\delta_\gamma} Q^{\text{coar}(\gamma)}_\bot \}_{\gamma \in \Gamma}$.
\end{definition}

Recall the definition of the pro $\Par[Q]$ from Remark \ref{remark:par}. Then as in Observation \ref{obs:free}, such assignments $\gamma \mapsto \delta_\gamma$ uniquely extend to pro morphisms $\delta : \mathscr{F}\Gamma \to \Par[Q]$, and we will also refer to such pro morphisms as deterministic monoidal automata. $\delta$ maps scalar string diagrams to one of the two functions $Q^0 \to Q^0_\bot$, and we use this to define the language of the automaton:

\begin{definition} \label{defn:lang_det}
  Let $\delta : \mathscr{F}\Gamma \to \Par[Q]$ be a deterministic monoidal automaton. Then the language accepted by $\delta$ is $\Lang(\delta) := \{ \alpha \in \mathscr{F}\Gamma(0, 0) \mid \delta_\alpha(\bullet) = \bullet \}$.
\end{definition}

We give a necessary condition for a monoidal language to be recognized by a deterministic monoidal automaton. The idea is to generalize the characterization of top-down deterministically recognizable tree languages as those that are closed under the operation of splitting a tree language into the set of possible paths through the trees, and reconstituting trees by grafting compatible paths \cite{steinby}. For string diagrams, we call the analogue of paths through a tree the \emph{causal histories} of a diagram (Definition \ref{defn:causal_hist}).

First, we briefly recall the machinery of (cartesian) \emph{restriction categories} \cite{COCKETT2002223}, that will be necessary in the following. Restriction categories are an abstraction of the category of partial functions, and provide us with a diagrammatic calculus for reasoning about determinization of monoidal languages.

\begin{definition}[\cite{cockett_lack_2007}] \label{defn:crcat}
  A cartesian restriction prop is a prop in which every object is equipped with a commutative comonoid structure (with the counit depicted by \raisebox{.25em}{\begin{tikzpicture}[xscale=.5,baseline=(current bounding box.center)]
\counit
\end{tikzpicture}}, comultiplication by \raisebox{.25em}{\begin{tikzpicture}[scale=.75,baseline=(current bounding box.center)]
\comult
\end{tikzpicture}
}, and symmetry by \raisebox{.25em}{\begin{tikzpicture}[scale=.3]
	\begin{pgfonlayer}{nodelayer}
		\node [style=none] (0) at (-0.75, -0.5) {};
		\node [style=none] (1) at (-0.75, 0.5) {};
		\node [style=none] (2) at (0.75, -0.5) {};
		\node [style=none] (3) at (0.75, 0.5) {};
	\end{pgfonlayer}
	\begin{pgfonlayer}{edgelayer}
		\draw [in=180, out=0] (1.center) to (2.center);
		\draw [in=180, out=0] (0.center) to (3.center);
	\end{pgfonlayer}
\end{tikzpicture}
}) that is coherent, and for which the comultiplication is natural (see Appendix \ref{appendix:crc} for details).
\end{definition}

\begin{definition}
  The free cartesian restriction prop on a monoidal graph $\mathcal{M}$, denoted $\mathscr{F}_{\downarrow}\mathcal{M}$ is given by taking the free prop on the monoidal graph $\mathcal{M}$ extended with a comultiplication and counit generator for every object in $V_{\mathcal{M}}$, and quotienting the morphisms by the structural equations of cartesian restriction categories (Appendix \ref{appendix:crc}).
\end{definition}

\begin{remark}
  $\Par$ is the paradigmatic example of a cartesian restriction category, with \raisebox{.25em}{\begin{tikzpicture}[xscale=.5,baseline=(current bounding box.center)]
\counit
\end{tikzpicture}} on $X$ given by the relation $X \to \{\bullet, \bot\}$ sending every element to $\bullet$, and \raisebox{.25em}{\begin{tikzpicture}[scale=.75,baseline=(current bounding box.center)]
\comult
\end{tikzpicture}
} given by the diagonal relation. $\Par_Q$ inherits this structure and so is a cartesian restriction prop. Therefore deterministic monoidal automata $(Q,\delta_\Gamma)$ also have inductive extensions to morphisms of cartesian restriction props, $\overline{\delta} : \mathscr{F}_{\downarrow}\Gamma \to \Par[Q]$, and these have a obvious notion of associated language, defined similarly to Definition \ref{defn:lang_det}. These are related by the following lemma, which follows from the universal properties of $\mathscr{F}\Gamma$ and $\mathscr{F}_{\downarrow}\Gamma$:
\end{remark}

\begin{lemma}
  If $(Q,\delta_\Gamma)$ is a deterministic monoidal automaton, then $\delta$ factors through $\overline{\delta}$ as $\delta = \mathscr{H}_\Gamma \comp \overline{\delta}$, where $\mathcal{H}_\Gamma : \mathscr{F}\Gamma \to \mathscr{F}_{\downarrow}\Gamma$ sends morphisms to their equivalence class in $\mathscr{F}_{\downarrow}\Gamma$.
\end{lemma}

Recall that any restriction category is poset-enriched: $f \leqslant g$ if $f$ is ``less defined'' than $g$, i.e. if $f$ coincides with $g$ on $f$'s domain of definition. For the hom-set from the monoidal unit to itself, we have $f \leqslant g \iff f \otimes g = f$. Now we can define causal histories:

\begin{definition} \label{defn:causal_hist}
  Let $\gamma$ be a string diagram in $\mathscr{F}\Gamma(0,0)$. We call a string diagram $h$ in $\mathscr{F}_{\downarrow}\Gamma(0,0)$ a causal history of $\gamma$ if $\mathcal{H}_\Gamma(\gamma) \leqslant h$ in $\mathcal{F}_{\downarrow}\Gamma(0,0)$. Let $L \subseteq \mathscr{F}\Gamma(0,0)$ be a regular monoidal language. The set of causal histories of $L$, denoted $\text{ch}(L)$, is defined to be $\mathcal{H}_\Gamma(L)^{\uparrow}$, the upwards closure of $\mathcal{H}_\Gamma(L)$ in the poset $\mathscr{F}_{\downarrow}\Gamma(0,0)$.
\end{definition}

A causal history represents the possible causal influence of parts of a diagram on generators appearing ``later'' in the diagram. For example, the following five string diagrams are causal histories of the rightmost string diagram below (every diagram is a causal history of itself), taken from the language introduced in Example \ref{ex:nondet}:

\begin{center}
  \includegraphics[scale=0.5]{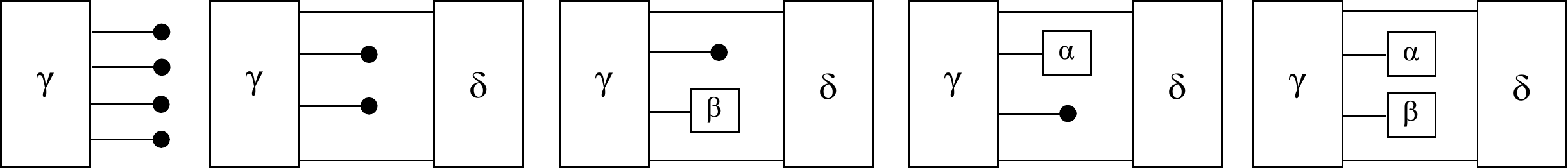}
\end{center}

\begin{lemma} \label{lemma:acc_chist}
  Let $M = (Q, \delta_\Gamma)$ be a deterministic monoidal automaton, with functors $\delta : \mathscr{F}\Gamma \to \Par[Q]$, $\overline{\delta} : \mathscr{F}_{\downarrow}\Gamma \to \Par[Q]$. Then if $\delta$ accepts $\gamma$, $\overline{\delta}$ accepts all causal histories of $\gamma$.
\end{lemma}
\begin{proof}
  Since $\delta =  \mathcal{H}_\Gamma \comp \overline{\delta}$, if $\delta$ accepts $\gamma$, then $\overline{\delta}$ accepts $\mathcal{H}_\Gamma(\gamma)$. Let $h$ be a causal history of $\gamma$. Then $\overline{\delta}(\mathcal{H}_\Gamma(\gamma)) = \overline{\delta}(h \otimes \mathcal{H}_\Gamma(\gamma)) = \overline{\delta}(h) \otimes \overline{\delta}(\mathcal{H}_\Gamma(\gamma))$. But then $\overline{\delta}$ accepts $h$ by Lemma \ref{lemma:closed_monoidal_product}.
\end{proof}

\begin{definition}[Causal closure of a language] \label{defn:closure}
Let L be a monoidal language over a monoidal alphabet $\Gamma$. Let $\bigotimes\!\text{ch}(L)$ denote the closure of the set of causal histories of $L$ under monoidal product. Then the causal closure of L is $\mathcal{H}_\Gamma^{-1}\bigotimes\!\text{ch}(L)$. A monoidal language is causally closed if it is equal to its causal closure.
\end{definition}

To illustrate causal closure, consider the following figure, which shows part of the derivation of a morphism in the causal closure of the language of Example \ref{ex:nondet}:
\begin{center}
  \resizebox{\width}{2.5cm}{\includegraphics[scale=0.5]{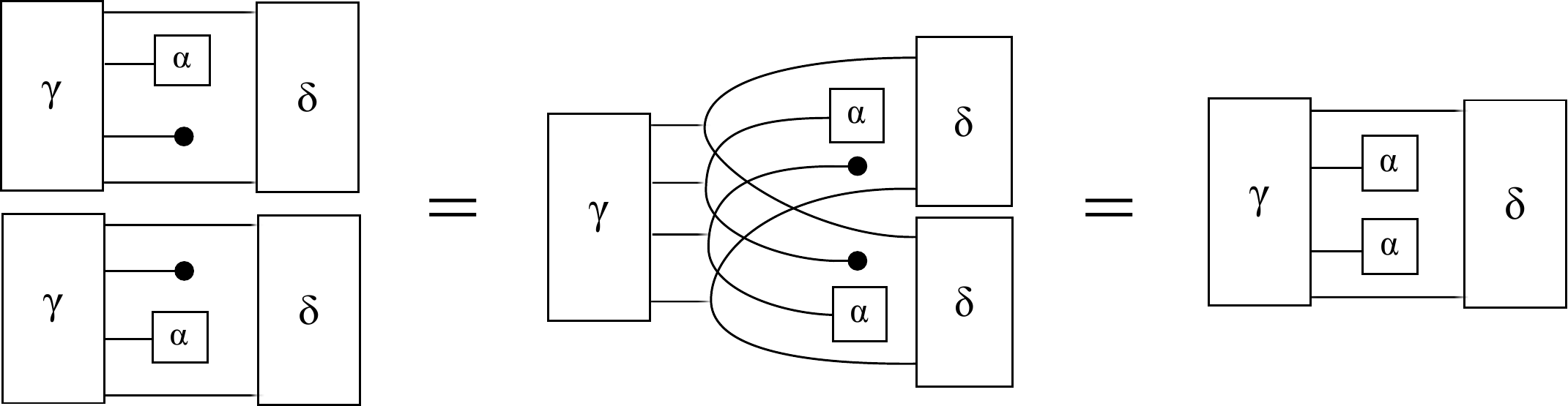}}
\end{center}

The leftmost diagram depicts the monoidal product of two causal histories determined by the counterexample language. By the equational theory of cartesian restriction categories (Appendix \ref{appendix:crc}), this is equal to the string diagrams in the center and on the right, where we first apply the naturality of \raisebox{.25em}{\begin{tikzpicture}[scale=.75,baseline=(current bounding box.center)]
\comult
\end{tikzpicture}
} (for $\gamma$), then unitality (twice), then naturality of \raisebox{.25em}{\begin{tikzpicture}[scale=.75,baseline=(current bounding box.center)]
\comult
\end{tikzpicture}
} (for $\delta$). The rightmost form of the diagram exhibits this morphism as being in the image of $\mathcal{H}_\Gamma$, and its preimage under $\mathcal{H}_\Gamma$ is the same diagram in $\mathscr{F}\Gamma$. Since this diagram is not in the original language, the language is not causally closed.

\begin{theorem} \label{thm:det_lang}
  If a monoidal language is recognized by a deterministic monoidal automaton, then it is causally closed.
\end{theorem}
\begin{proof}
  Let $L$ be recognized by a deterministic monoidal automaton $\delta : \mathscr{F}\Gamma \to \Par[Q]$. We have $\delta = \mathcal{H}_\Gamma \comp  \overline{\delta}$ and from Lemma \ref{lemma:acc_chist} that $\overline{\delta}$ accepts causal histories of morphisms in $L$. Since languages are closed under monoidal product (Lemma \ref{lemma:closed_monoidal_product}), then by definition of the causal closure, $\delta$ must accept everything in the causal closure of $L$.
\end{proof}

\section{Deterministically recognizable monoidal languages} \label{sec:determinizability}

Non-deterministic finite state automata for words and bottom-up trees can be determinized via the well known powerset construction. However, top-down tree automata cannot be determinized in general \cite[\S 2.11]{steinby}, so general monoidal automata also cannot be determinized (Observation \ref{ex:tree_aut}). However, there are interesting examples of deterministically recognizable monoidal languages that are not tree languages, such as the monoidal Dyck language (Example \ref{ex:brackets}) and Sierpi\'nski gaskets (Example \ref{ex:sierpinski}), and it is an intriguing theoretical challenge to characterize such languages.

In Section \ref{sec:determinization} we study a class of determinizable automata called \emph{convex} automata. In Section \ref{sec:sufficient} we give a sufficient condition for a language to be deterministically recognizable.

\subsection{Convex automata and the powerset construction} \label{sec:determinization}
The classical powerset construction is given conceptually by composition with the functor $\Rel \to \textsf{Set}$, right adjoint to the inclusion $\textsf{Set} \hookrightarrow \Rel$. As remarked above, we cannot hope to obtain an analogue of this functor for monoidal automata. Thus we describe a suitable subcategory of $\Rel[Q]$ for which determinization is functorial, that of convex relations.

\begin{definition}
  A relation $\Delta : Q^n \to \mathscr{P}(Q^m)$ is convex if there is a morphism $\Delta^{\!*}$ such that the following square commutes:
\[\begin{tikzcd}[ampersand replacement=\&,every label/.append style = {font = \small}]
	{(\mathscr{P}Q)^n} \&\& {(\mathscr{P}Q)^m} \\
	{\mathscr{P}(Q^n)} \&\& {\mathscr{P}(Q^m)}
	\arrow["{\Delta^{\!*}}", from=1-1, to=1-3]
	\arrow["{\Delta^{\#}}", from=2-1, to=2-3]
	\arrow["\nabla"', from=1-1, to=2-1]
	\arrow["\nabla", from=1-3, to=2-3]
\end{tikzcd}\]

where $\Delta^{\#}$ is the Kleisli lift of $\Delta$, and $\nabla$ is the monoidal multiplication given by the commutativity of the powerset monad.
\end{definition}

\begin{observation} \label{remark:star_unique}
If $\Delta$ is convex, the morphism $\Delta^{\!*}$ is unique, since $\nabla$ is a monomorphism.
\end{observation}

\begin{example}
The relation $\Delta_\gamma : Q^0 \to \mathscr{P}(Q^4)$ induced by the grammar in Example \ref{ex:nondet} is not convex, since $(A,B,B,A)$ and $(A,C,C,A)$, which we can think of as ``convex combinations'' of the other state vectors, are not included in the image of the relation.
\end{example}

\begin{lemma} \label{lemma:crel_is_sub_pro}
Convex relations determine a sub-pro $\textsf{CRel}_Q \hookrightarrow \textsf{Rel}_Q$.
\end{lemma}
\begin{proof}
See Appendix \ref{appendix:subpro}.
\end{proof}

\begin{definition}
An automaton $\Delta : \mathscr{F}\Gamma \to \textsf{Rel}_Q$ is convex if it factors through $\textsf{CRel}_Q$.
\end{definition}

The following lemma gives the powerset construction on convex automata. We use the non-empty powerset $\mathscr{P}^{+}$ to avoid duplication of failure state ($\varnothing$ in $\Rel[Q]$, but $\bot$ in $\Par[\mathscr{P}^{+}(Q)]$):

\begin{lemma} \label{lemma:detr}
  For each set Q there is a morphism of pros $\detr_Q : \CRel[Q] \to \Par[\mathscr{P}^{+}(Q)]$ which is identity on objects and acts as follows on morphisms:
  \[\begin{tikzcd}[ampersand replacement=\&,row sep=3mm]
	\Delta_\alpha : Q^n \to \mathscr{P}(Q^m)  \\
	\mathscr{P}^{+}(Q)^n \xrightarrow[]{\eta^n} (\bot\mathscr{P}^{+}(Q))^n \xrightarrow[]{\cong} \mathscr{P}(Q)^n \xrightarrow[]{\Delta_{\alpha}^{\!*}} \mathscr{P}(Q)^m \xrightarrow[]{\cong} (\bot\mathscr{P}^{+}(Q))^m \xrightarrow[]{\nabla} \bot\mathscr{P}^{+}(Q)^m
	\arrow[maps to, from=1-1, to=2-1]
      \end{tikzcd}\]

    where $\bot$ is the maybe monad, $\eta$ is the unit of this monad, and $\nabla$ is its monoidal multiplication with respect to the cartesian product.
\end{lemma}
\begin{proof}
  See Appendix \ref{sec:lang_preserved}.
\end{proof}

Determinization of a convex automaton $\Delta : \mathscr{F}\Gamma \to \CRel[Q]$ is now just given by post-composition with the functor $\detr_Q$. We show that this preserves the language:
\begin{theorem}
  Determinization of convex automata preserves the accepted language: let $\Delta : \mathscr{F}\Gamma \to \textsf{CRel}_Q$ be a convex automaton, then $\Lang(\Delta) = \Lang(\Delta \comp \detr_Q)$.
\end{theorem}
\begin{proof}
  Let $\alpha \in \mathscr{L}(\Delta)$, i.e. $\Delta_\alpha(\bullet) = \{ \bullet \}$. Then we must have $\Delta_\alpha^{*}(\bullet) = \bullet$, and so $(\Delta \comp \detr_Q)_\alpha(\bullet) = \bullet$. Conversely let $\alpha \in \mathscr{L}_{\detr}(\Delta \comp \detr_Q)$, i.e. $(\Delta \comp \detr_Q)_\alpha(\bullet) = \bullet$. Then we must have that $\Delta_{\alpha}^{*}(\bullet) = \bullet$, and so $\Delta_{\alpha}(\bullet) = \{\bullet\}$, that is $\alpha \in \Lang(\Delta)$.
\end{proof}

\begin{example}
  Non-deterministic monoidal automata over word monoidal alphabets (Definition \ref{defn:word_alphabet}) are convex: for a relation $\Delta : Q \to \mathscr{P}(Q)$, $\Delta^{\!*}$ is given by the Kleisli extension of $\Delta$. This reflects the well known determinizability of classical finite-state automata.
\end{example}

\begin{example}
  Similarly, non-deterministic monoidal automata over bottom-up tree monoidal alphabets (Definition \ref{defn:tree_alphabet}) are convex, with $\Delta^{\!*} := \nabla \fatsemi \Delta^{\#}$. For top-down tree monoidal alphabets, the general obstruction to convexity (and thus determinizability) is seen as the non-existence of a left inverse of $\nabla$.
\end{example}

\subsection{A sufficient condition for deterministic recognizability} \label{sec:sufficient}
\begin{theorem} \label{thm:sufficient}
  If the syntactic pro of a regular monoidal language has the structure of a cartesian restriction prop, then the language is recognizable by a deterministic monoidal automaton.
\end{theorem}
\begin{proof}
  Let $L$ be a monoidal language such that $\mathscr{F}\Gamma/{\equiv}_L$ has a cartesian restriction prop structure. We exhibit a pro morphism $\mathscr{F}\Gamma/{\equiv_L} \xrightarrow[]{\phi} \Par[Q]$ such that $\mathscr{F}\Gamma \xrightarrow[]{S_L} \mathscr{F}\Gamma/{\equiv}_L \xrightarrow[]{\phi} \Par[Q]$ is a deterministic monoidal automaton accepting exactly $L$.

  Let $Q := \mathscr{F}\Gamma/{\equiv}_L(0,1)$. By Lemma \ref{lemma:reg_then_lf}, this is a finite set. For $m > 0$ and $\left[ \beta \right] \in \mathscr{F}\Gamma/{\equiv_L}(n,m)$, define $\phi(\left[ \beta \right]) : n \to m$ to be the following map from $Q^n \to Q^m_\bot$:
  \begin{center}
  \resizebox{\width}{2.2cm}{\includegraphics[scale=0.75]{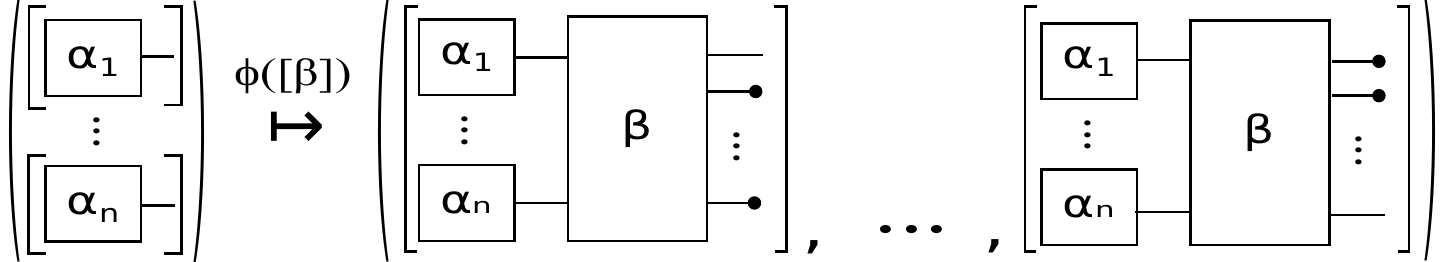}}
  \end{center}
  When $m = 0$ (i.e. $\left[\beta\right]$ has coarity 0), let $\phi(\left[\beta\right])([\alpha_1], ..., [\alpha_n]) = \bullet$, if $\left[ (\alpha_1 \otimes ... \otimes \alpha_n) \fatsemi \beta \right] = \left[ \raisebox{.25em}{\begin{tikzpicture}[scale=.5,baseline=(current bounding box.center)]
\akasa
\end{tikzpicture}}\!\! \right]$, and $\phi(\left[\beta\right])([\alpha_1], ..., [\alpha_n])=\bot$ otherwise. The proof that this defines a morphism of pros is an exercise in diagrammatic reasoning using the equational theory of cartesian restriction categories and is left to Appendix \ref{appendix:sufficient}. To see that this automaton accepts exactly $L$, let $\alpha \in \Lang(S_L \comp \phi)$, then by definition we must have $S_L(\alpha) = \left[ \raisebox{.25em}{\begin{tikzpicture}[scale=.5,baseline=(current bounding box.center)]
\akasa
\end{tikzpicture}}\!\! \right]$, and so $\alpha \in L$ (by Lemma \ref{lemma:recog_via_empty}). Conversely let $\alpha \in L$, then $S_L(\alpha) = \left[ \raisebox{.25em}{\begin{tikzpicture}[scale=.5,baseline=(current bounding box.center)]
\akasa
\end{tikzpicture}} \!\! \right]$ and by definition $\phi\left(\left[ \raisebox{.25em}{\begin{tikzpicture}[scale=.5,baseline=(current bounding box.center)]
\akasa
\end{tikzpicture}}\!\!\right]\right)(\bullet) = \bullet$, so $\alpha \in \Lang(S_L \comp \phi)$. Therefore $S_L\comp\phi$ is a deterministic monoidal automaton recognizing $L$.
\end{proof}

\begin{example}
  A simple example is given by the language $L$ of ``bones'' over the monoidal alphabet $\Gamma = \{ \!\raisebox{.25em}{\begin{tikzpicture}[xscale=.5,baseline=(current bounding box.center)]
\unit
\end{tikzpicture}}\!, ~\raisebox{.25em}{\begin{tikzpicture}[xscale=.5,baseline=(current bounding box.center)]
\counit
\end{tikzpicture}} \}$, having one connected component: \begin{tikzpicture}[xscale=.5,baseline=(current bounding box.center)]
\draw[draw=none] (0,0) rectangle (1,0.4);
\node[dot] at (0*\len,1.2*\len) {};
\draw[wire] (0*\len,1.2*\len) -- (1,1.2*\len);
\node[dot] at (1,1.2*\len) {};
\end{tikzpicture}
. The syntactic pro of this language has a cartesian restriction prop structure, with the counit given by the equivalence class $\left[\raisebox{.25em}{\begin{tikzpicture}[xscale=.5,baseline=(current bounding box.center)]
\counit
\end{tikzpicture}}\!\right]$, comultiplication by $[\begin{tikzpicture}[xscale=.5,baseline=(current bounding box.center)]
\draw[draw=none] (0,0) rectangle (1,0.5);

\draw[wire] (2*\len,1.5*\len) -- (0,1.5*\len);
\node[dot] at (2*\len,1.5*\len) {};

\draw[wire] (2*\len,2*\len) -- (1,2*\len);
\node[dot] at (2*\len,2*\len) {};

\draw[wire] (2*\len,1*\len) -- (1,1*\len);
\node[dot] at (2*\len,1*\len) {};
\end{tikzpicture}
]$, and symmetry by $[\begin{tikzpicture}[xscale=.5,baseline=(current bounding box.center)]
\draw[draw=none] (0,0) rectangle (1,0.4);

\draw[wire] (2*\len,2*\len) -- (0,2*\len);
\node[dot] at (2*\len,2*\len) {};

\draw[wire] (2*\len,1*\len) -- (0,1*\len);
\node[dot] at (2*\len,1*\len) {};

\draw[wire] (3*\len,2*\len) -- (1.25,2*\len);
\node[dot] at (3*\len,2*\len) {};

\draw[wire] (3*\len,1*\len) -- (1.25,1*\len);
\node[dot] at (3*\len,1*\len) {};
\end{tikzpicture}]$. It is clear that $\mathscr{F}\Gamma/{\equiv_L}(0,1)$ has one equivalence class, $\left[\raisebox{.25em}{\begin{tikzpicture}[xscale=.5,baseline=(current bounding box.center)]
\counit
\end{tikzpicture}}\!\right]$, which becomes the state of the monoidal automaton. The construction above then gives the obvious transition functions required for each generator.

\end{example}

\section{Conclusion and future work}
The most immediate open question is to determine necessary and sufficient conditions for determinizability: causal closure is a promising candidate. Furthermore we would like to understand the relation between convexity and Theorem \ref{thm:sufficient}. Classical topics in the theory of regular languages such as a Myhill-Nerode theorem are also ripe for future investigation. We also plan to investigate further applications of regular monoidal languages in computer science, for example representing trace languages and look-ahead parsing.

Just as our definition of regular monoidal grammar was obtained from Walters' definition of regular grammar by replacing the adjunction $\textsf{Cat} \to \textsf{Graph}$ with the adjunction $\textsf{Pro} \to \textsf{MonGraph}$, we might consider other adjunctions and their corresponding notion of grammar. In the first instance, our theory should smoothly generalize to languages in free props, but perhaps also other (higher) categorical structures.

We plan to investigate a notion of context-free monoidal language, using a similar algebraic approach to this paper. One candidate for the algebra of such languages, inspired again by \cite{WALTERS1989199}, are (monoidal) multicategories of $n$-hole contexts (in the sense of Definition \ref{defn:context}).

\bibliography{main}

\begin{thebibliography}{10}

\bibitem{mmso}
Miko{\l}aj Boja{\'{n}}czyk, Bartek Klin, and Julian Salamanca.
\newblock Monadic monadic second order logic.
\newblock 2022.
\newblock \href {http://arxiv.org/abs/arXiv:2201.09969}
  {\path{arXiv:arXiv:2201.09969}}.

\bibitem{bossut}
Francis Bossut, Max Dauchet, and Bruno Warin.
\newblock A {Kleene} theorem for a class of planar acyclic graphs.
\newblock {\em Inf. Comput.}, 117:251--265, 03 1995.
\newblock \href {https://doi.org/10.1006/inco.1995.1043}
  {\path{doi:10.1006/inco.1995.1043}}.

\bibitem{COCKETT2002223}
J.R.B. Cockett and Stephen Lack.
\newblock Restriction categories {I}: categories of partial maps.
\newblock {\em Theoretical Computer Science}, 270(1):223--259, 2002.
\newblock \href {https://doi.org/https://doi.org/10.1016/S0304-3975(00)00382-0}
  {\path{doi:https://doi.org/10.1016/S0304-3975(00)00382-0}}.

\bibitem{cockett_lack_2007}
Robin Cockett and Stephen Lack.
\newblock Restriction categories {III}: colimits, partial limits and
  extensivity.
\newblock {\em Mathematical Structures in Computer Science}, 17(4):775–817,
  2007.
\newblock \href {https://doi.org/10.1017/S0960129507006056}
  {\path{doi:10.1017/S0960129507006056}}.

\bibitem{lmcs:6213}
Thomas Colcombet and Daniela Petrişan.
\newblock {Automata Minimization: a Functorial Approach}.
\newblock {\em {Logical Methods in Computer Science}}, {Volume 16, Issue 1},
  March 2020.
\newblock URL: \url{https://lmcs.episciences.org/6213}, \href
  {https://doi.org/10.23638/LMCS-16(1:32)2020}
  {\path{doi:10.23638/LMCS-16(1:32)2020}}.

\bibitem{di2021functorial}
Ivan Di~Liberti, Fosco Loregian, Chad Nester, and Pawe{\l} Soboci{\'n}ski.
\newblock Functorial semantics for partial theories.
\newblock {\em Proceedings of the ACM on Programming Languages}, 5(POPL):1--28,
  2021.

\bibitem{fahrenberg}
Uli Fahrenberg, Christian Johansen, Georg Struth, and Krzysztof Ziemiański.
\newblock Languages of higher-dimensional automata.
\newblock {\em Mathematical Structures in Computer Science}, 31(5):575–613,
  2021.
\newblock \href {https://doi.org/10.1017/S0960129521000293}
  {\path{doi:10.1017/S0960129521000293}}.

\bibitem{10.1093/logcom/exx029}
Richard Garner and Tom Hirschowitz.
\newblock {Shapely monads and analytic functors}.
\newblock {\em Journal of Logic and Computation}, 28(1):33--83, 11 2017.
\newblock \href {https://doi.org/10.1093/logcom/exx029}
  {\path{doi:10.1093/logcom/exx029}}.

\bibitem{steinby}
Ferenc Gécseg and Magnus Steinby.
\newblock Tree automata, 2015.
\newblock \href {https://doi.org/10.48550/ARXIV.1509.06233}
  {\path{doi:10.48550/ARXIV.1509.06233}}.

\bibitem{Heindel2017AMT}
T.~Heindel.
\newblock A {Myhill-Nerode} theorem beyond trees and forests via finite
  syntactic categories internal to monoids.
\newblock {\em Preprint}, 2017.

\bibitem{power_robinson_1997}
John Power and Edmund Robinson.
\newblock Premonoidal categories and notions of computation.
\newblock {\em Mathematical Structures in Computer Science}, 7(5), 1997.
\newblock \href {https://doi.org/10.1017/S0960129597002375}
  {\path{doi:10.1017/S0960129597002375}}.

\bibitem{Rosenthal1995}
Kimmo~I. Rosenthal.
\newblock Quantaloids, enriched categories and automata theory.
\newblock {\em Applied Categorical Structures}, 3(3):279--301, 1995.
\newblock \href {https://doi.org/10.1007/bf00878445}
  {\path{doi:10.1007/bf00878445}}.

\bibitem{10.1371/journal.pbio.0020424}
Paul W.~K Rothemund, Nick Papadakis, and Erik Winfree.
\newblock Algorithmic self-assembly of {DNA} {S}ierpinski triangles.
\newblock {\em PLOS Biology}, 2(12), 12 2004.
\newblock \href {https://doi.org/10.1371/journal.pbio.0020424}
  {\path{doi:10.1371/journal.pbio.0020424}}.

\bibitem{sakarovitch2009elements}
Jacques Sakarovitch.
\newblock {\em Elements of automata theory}.
\newblock Cambridge University Press, Cambridge New York, 2009.

\bibitem{Selinger2011}
P.~Selinger.
\newblock A survey of graphical languages for monoidal categories.
\newblock In B.~Coecke, editor, {\em New Structures for Physics}, pages
  289--355. Springer Berlin Heidelberg, Berlin, Heidelberg, 2011.
\newblock \href {https://doi.org/10.1007/978-3-642-12821-9_4}
  {\path{doi:10.1007/978-3-642-12821-9_4}}.

\bibitem{eilenbergforfree}
Henning Urbat, Jiri Ad{\'a}mek, Liang-Ting Chen, and Stefan Milius.
\newblock {Eilenberg Theorems for Free}.
\newblock In Kim~G. Larsen, Hans~L. Bodlaender, and Jean-Francois Raskin,
  editors, {\em 42nd International Symposium on Mathematical Foundations of
  Computer Science (MFCS 2017)}, volume~83 of {\em LIPIcs}, pages 43:1--43:15,
  Dagstuhl, Germany, 2017.
\newblock \href {https://doi.org/10.4230/LIPIcs.MFCS.2017.43}
  {\path{doi:10.4230/LIPIcs.MFCS.2017.43}}.

\bibitem{WALTERS1989199}
R.F.C. Walters.
\newblock A note on context-free languages.
\newblock {\em Journal of Pure and Applied Algebra}, 62(2):199--203, 1989.
\newblock \href {https://doi.org/10.1016/0022-4049(89)90151-5}
  {\path{doi:10.1016/0022-4049(89)90151-5}}.

\bibitem{zamdzhiev2016}
Vladimir Zamdzhiev.
\newblock {\em Rewriting Context-free Families of String Diagrams}.
\newblock PhD thesis, University of Oxford, 2016.

\end{thebibliography}

\appendix

\section{String diagrams for monoidal categories} \label{appendix:string}
We briefly recall the string diagram notation for morphisms in free (strict) monoidal categories. Recall that a strict monoidal category is a category $\mathscr{C}$ equipped with a functor $\otimes : \mathscr{C} \times \mathscr{C} \to \mathscr{C}$ (monoidal product), and an object $I \in \mathscr{C}$ (monoidal unit), such that for all objects $A,B,C \in \mathscr{C}$, $A \otimes (B \otimes C) = (A \otimes B) \otimes C$ and $A \otimes I = I = I \otimes A$.

We choose a left to right convention for our string diagrams. Objects are depicted as wires labelled by the object, and by convention the monoidal unit is not drawn, i.e. it is represented by the empty diagram \raisebox{.25em}{\begin{tikzpicture}[scale=.5,baseline=(current bounding box.center)]
\akasa
\end{tikzpicture}}. The monoidal product of objects is depicted by wires drawn in parallel. A morphism $f : A_1 \otimes ... \otimes A_n \to B_1 \otimes ... \otimes B_m$ is depicted as a box labelled by $f$ with wires $A_1 \otimes ... \otimes A_n$ entering on the left and $B_1 \otimes ... \otimes B_m$ exiting on the right. By convention, no box is drawn for identity morphisms, and so the identity on the monoidal unit is also the empty diagram. The monoidal product $f \otimes g$ of morphisms $f : A_1 \otimes ... \otimes A_n \to B_1 \otimes ... \otimes B_m, g : C_1 \otimes ... \otimes C_{n'} \to D_1 \otimes ... \otimes D_{m'}$ is depicted by writing the morphisms in parallel. Sequential composition $f \comp h$ of morphisms $f : A_1 \otimes ... \otimes A_n \to B_1 \otimes ... \otimes B_m, h : B_1 \otimes ... \otimes B_m \to E_1 \otimes ... \otimes E_p$ is depicted by joining the outgoing wires of one box to the incoming wires of another. In summary we have:

\begin{center}
  \includegraphics[scale=0.7]{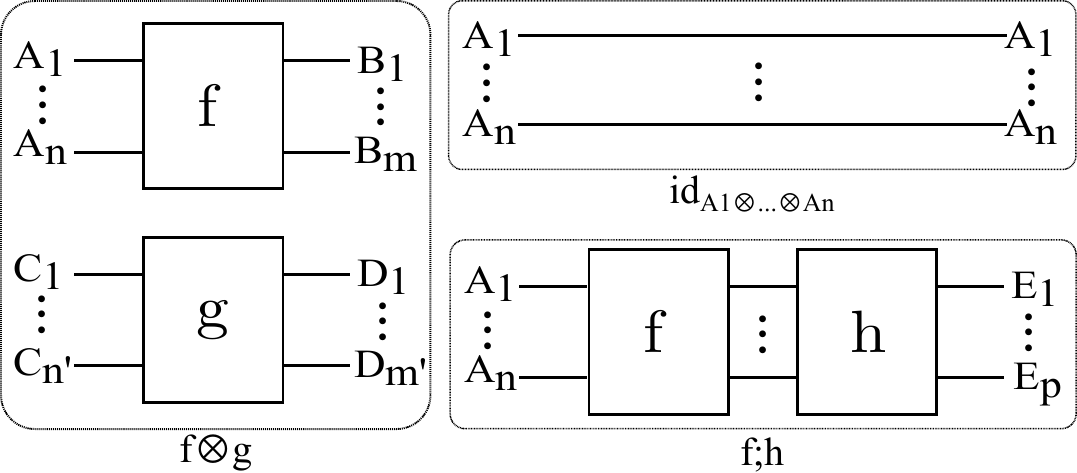}
\end{center}

This notation is sound and complete: an equation between morphisms of strict monoidal categories follows from the axioms of strict monoidal categories if and only if it holds between string diagrams up to planar isotopy. Working with string diagrams rather than the usual term syntax for morphisms is more intuitive, and leads to shorter proofs, since the structural equations (such as associativity of composition and monoidal product) hold automatically. See \cite[\S 3]{Selinger2011} for more details and further references.

\section{Congruence on a monoidal category} \label{appendix:congruence}

\begin{definition} \label{defn:cong}
  A congruence on a monoidal category $\mathscr{C}$ is an equivalence relation $f\sim g$ on pairs of parallel morphisms $f,g : x \to x'$ compatible with composition and monoidal product:
  \begin{itemize}
  \item $f \sim g \implies k \circ f \circ h \sim k \circ g \circ h$ ~~~ whenever these composites are defined
  \item $f \sim g \implies p \otimes f \otimes q \sim p \otimes g \otimes q$
  \end{itemize}
\end{definition}

Given a congruence on a monoidal category $\mathscr{C}$, we can define the quotient monoidal category $\mathscr{C}/{\sim}$ as the category with objects those of $\mathscr{C}$, and homsets $(\mathscr{C}/{\sim})(x,x') \coloneqq \mathscr{C}(x,x')/{\sim}$, with composition and monoidal product defined in the obvious way. The quotient functor $\mathscr{C} \to \mathscr{C}/{\sim}$ is monoidal, full and bijective on objects (and strict when $\mathscr{C}$ is strict). When $\mathscr{C}$ is a pro, the quotient monoidal category is a pro, and the quotient functor is a pro morphism. One can easily verify with string diagrams that the syntactic congruence (Definition \ref{defn:syntactic_cong}) is indeed a congruence, and so the syntactic pro is well defined.

\section{Cartesian restriction categories}  \label{appendix:crc}

A cartesian restriction category can be defined as a symmetric monoidal category in which every object is equipped with a coherent commutative comonoid structure for which comultiplication is natural. The following equations spell out the details of this definition. We write \raisebox{.25em}{\begin{tikzpicture}[xscale=.5,baseline=(current bounding box.center)]
\counit
\end{tikzpicture}} for the counit of the comonoid on an arbitrary object, \raisebox{.25em}{\begin{tikzpicture}[scale=.75,baseline=(current bounding box.center)]
\comult
\end{tikzpicture}
} for the comultiplication of the comonoid on an arbitrary object, and \raisebox{.25em}{\begin{tikzpicture}[scale=.3]
	\begin{pgfonlayer}{nodelayer}
		\node [style=none] (0) at (-0.75, -0.5) {};
		\node [style=none] (1) at (-0.75, 0.5) {};
		\node [style=none] (2) at (0.75, -0.5) {};
		\node [style=none] (3) at (0.75, 0.5) {};
	\end{pgfonlayer}
	\begin{pgfonlayer}{edgelayer}
		\draw [in=180, out=0] (1.center) to (2.center);
		\draw [in=180, out=0] (0.center) to (3.center);
	\end{pgfonlayer}
\end{tikzpicture}
} for the symmetry between two objects. Then to say that there is a commutative comonoid structure on each object is to say that the following equations of string diagrams (CCM) hold (respectively: coassociativity, commutativity, and left unitality):
\begin{equation}\label{eq:comonoid}\tag{CCM}
                \raisebox{.5em}{\begin{tikzpicture}[xscale=-.75,baseline=(current bounding box.center)]
                                \umult
                                \lid
                                \step{\mult}
                                \ezs{\lmult
                                        \uid
                                        \step{\mult}}{xshift=3cm,yshift=\len cm}
                                \node at (2.5,2*\len) {$=$};
                        \end{tikzpicture}\hspace{2em}
                        \begin{tikzpicture}[xscale=-.75,baseline=(current bounding box.center)]
                                \braid\step{\mult}\ezs{\mult}{xshift=3cm}
                                \node at (2.5,2*\len) {$=$};
                        \end{tikzpicture}\hspace{2em}
                        \begin{tikzpicture}[xscale=-.75,baseline=(current bounding box.center)]
                                \ezs{\unit}{yshift=\len cm}\lid
                                \ezs{\id}{xshift=3cm}
                                \step{\mult}
                                \node at (2.5,2*\len) {$=$};
                        \end{tikzpicture}}
        \end{equation}

      Note that ``right unitality'' may be derived from these. To say that these comonoid structures are coherent is to say that for all objects X and Y we have the following equations of string diagrams:
  \begin{equation}\label{eq:coherent}\tag{coherent}
        \begin{tikzpicture}[xscale=.75,baseline=(current bounding box.center)]
                \comult
                \node[left] at (0,2*\len) {$\scriptscriptstyle X\otimes Y$};
                \node[right] at (1,\len) {$\scriptscriptstyle X\otimes Y$};
                \node[right] at (1,3*\len) {$\scriptscriptstyle X\otimes Y$};
                \ezs{
                        \comult
                        \ezs{\comult}{yshift=1cm}
                        \ezs{\braid}{xshift=1cm, yshift=.5cm}
                        \ezs{\lid}{xshift=1cm}
                        \ezs{\uid}{xshift=1cm, yshift=1cm}
                        \node[left] at (0,2*\len) {$\scriptscriptstyle X$};
                        \node[left] at (0,6*\len) {$\scriptscriptstyle Y$};
                        \node[right] at (2,\len) {$\scriptscriptstyle X$};
                        \node[right] at (2,5*\len) {$\scriptscriptstyle X$};
                        \node[right] at (2,3*\len) {$\scriptscriptstyle Y$};
                        \node[right] at (2,7*\len) {$\scriptscriptstyle Y$};
                }{xshift=3cm,yshift=-.5cm}
                \node at (2.25,2*\len) {\Scale[1.25]{=}};
        \end{tikzpicture}
        \qquad
        \begin{tikzpicture}[baseline=(current bounding box.center)]
                \counit
                \node[left] at (0,2*\len) {$\scriptscriptstyle X\otimes Y$};
                \ezs{
                        \ezs{\counit}{yshift=-\len cm}
                        \ezs{\counit}{yshift= \len cm}
                        \node[left] at (0,\len) {$\scriptscriptstyle X$};
                        \node[left] at (0,3*\len) {$\scriptscriptstyle Y$};
                }{xshift=2cm}
                \node at (1.225,2*\len) {\Scale[1.25]{=}};
\end{tikzpicture}
      \end{equation}

      Finally to say that comultiplication natural is to say that we can move morphisms through comultiplication as follows:
      \begin{equation}\label{eq:natural}\tag{natural}
        \begin{tikzpicture}[baseline=(current bounding box.center),xscale=.5]
                \mor{f}
                \ezs{\comult}{xshift=1cm}
                \node[left] at (0,2*\len) {$\scriptscriptstyle X$};
                \node[right] at (2,\len) {$\scriptscriptstyle Y$};
                \node[right] at (2,3*\len) {$\scriptscriptstyle Y$};
                \ezs{
                        \comult
                        \ezs{\umor{f}}{xshift=1cm}
                        \ezs{\lmor{f}}{xshift=1cm}
                        \node[left] at (0,2*\len) {$\scriptscriptstyle X$};
                        \node[right] at (2,\len) {$\scriptscriptstyle Y$};
                        \node[right] at (2,3*\len) {$\scriptscriptstyle Y$};
                }{xshift=5cm}
                \node at (3.5,2*\len)  {\Scale[1.25]{=}};
\end{tikzpicture}
        \end{equation}

\section{Details for Section \ref{sec:determinizability}}

\begin{proof}[Proof of Lemma \ref{lemma:crel_is_sub_pro}] \label{appendix:subpro}
  It is clear that identity relations are convex. It remains to show that the composite of convex relations is convex, and that the monoidal product of convex relations is convex. For the former, take convex relations $\Delta_\alpha : Q^a \to \mathscr{P}(Q^b), \Delta_\beta: Q^b \to \mathscr{P}(Q^c)$, and take $(\Delta_\beta \diamond \Delta_\alpha)^{*} = \Delta_\beta^{*} \circ \Delta_\alpha^{\!*}$, where $\diamond$ is composition in $\textsf{Kl}(\mathscr{P})$. Consider the following diagram:
\[\begin{tikzcd}[ampersand replacement=\&]
	{(\mathscr{P}Q)^a} \&\& {(\mathscr{P}Q)^b} \&\& {(\mathscr{P}Q)^c} \\
	{\mathscr{P}(Q^a)} \&\& {\mathscr{P}(Q^b)} \&\& {\mathscr{P}(Q^c)} \\
	\& {\mathscr{P}^2(Q^b)} \&\& {\mathscr{P}^2(Q^c)} \\
	\&\& {\mathscr{P}^3(Q^c)}
	\arrow["{\Delta_\alpha^{*}}", from=1-1, to=1-3]
	\arrow["{\Delta_\alpha^{\#}}", from=2-1, to=2-3]
	\arrow["\nabla"', from=1-1, to=2-1]
	\arrow["\nabla", from=1-3, to=2-3]
	\arrow["{\mathscr{P}(\Delta_\alpha)}"', from=2-1, to=3-2]
	\arrow["\mu"', from=3-2, to=2-3]
	\arrow["{\Delta_\beta^{*}}", from=1-3, to=1-5]
	\arrow["{\Delta_\beta^{\#}}", from=2-3, to=2-5]
	\arrow["\nabla", from=1-5, to=2-5]
	\arrow["{\mathscr{P}(\Delta_\beta)}"', from=2-3, to=3-4]
	\arrow["\mu"', from=3-4, to=2-5]
	\arrow["{\mathscr{P}(\mu)}"', from=4-3, to=3-4]
	\arrow["{\mathscr{P}^2(\Delta_\beta)}"', from=3-2, to=4-3]
\end{tikzcd}\]

We want to show that $\Delta_\beta^{\#} \circ \Delta_\alpha^{\#} = (\Delta_\beta \diamond \Delta_\alpha)^\#$, so that the pasting of the two convexity squares at the top witnesses convexity of the composite. By definition of Kleisli extension we have that: \begin{align*}
  &\Delta_\beta^{\#} \circ \Delta_\alpha^{\#} = \mu \circ \mathscr{P}(\Delta_\beta) \circ \mu \circ \mathscr{P}(\Delta_\alpha) \\
  \intertext{by naturality of $\mu$,}
  &= \mu \circ \mathscr{P}(\mu) \circ \mathscr{P}^2(\Delta_\beta) \circ \mathscr{P}(\Delta_\alpha) \\
  &= \mu \circ \mathscr{P}(\mu \circ \mathscr{P}(\Delta_\beta) \circ \Delta_\alpha) \\
  &= \mu \circ \mathscr{P}(\Delta_\beta \diamond \Delta_\alpha) \\
  &= (\Delta_\beta \diamond \Delta_\alpha)^{\#}
\end{align*}

Now take convex relations $\Delta_\gamma : Q^{n_1} \to \mathscr{P}(Q^{m_1}), \Delta_\varepsilon : Q^{n_2} \to \mathscr{P}(Q^{m_2})$. Take $(\Delta_\gamma \otimes \Delta_\varepsilon)^{*} = \Delta_\gamma^{*} \times \Delta_\varepsilon^{*}$. We have that:
    \begin{align*}
      &\mathscr{P}(Q)^{n_1+n_2} \xrightarrow[]{(\Delta_\gamma \otimes \Delta_\varepsilon)^{*}} \mathscr{P}(Q)^{m_1 + m_2} \xrightarrow[]{\nabla} \mathscr{P}(Q^{m_1 + m_2})  \\
      &= \mathscr{P}(Q)^{n_1+n_2} \xrightarrow[]{\langle \nabla \circ \Delta_\gamma^{*}, \nabla \circ \Delta_\varepsilon^{*}\rangle}  \mathscr{P}(Q^{m_1}) \times \mathscr{P}(Q^{m_2}) \xrightarrow[]{\nabla} \mathscr{P}(Q^{m_1 + m_2}) \\ \intertext{by convexity of $\Delta_\gamma, \Delta_\varepsilon$,}
      &= \mathscr{P}(Q)^{n_1+n_2} \xrightarrow[]{\nabla \times \nabla} \mathscr{P}(Q^{n_1}) \times \mathscr{P}(Q^{n_2}) \xrightarrow[]{\mathscr{P}(\Delta_\gamma) \times \mathscr{P}(\Delta_\varepsilon)} \mathscr{PP}(Q^{m_1}) \times \mathscr{PP}(Q^{m_2}) \\
      &  \hspace*{70mm}\xrightarrow[]{\mu \times \mu} \mathscr{P}(Q^{m_1}) \times \mathscr{P}(Q^{m_2}) \xrightarrow[]{\nabla} \mathscr{P}(Q^{m_1 + m_2}) \\
      &= \mathscr{P}(Q)^{n_1 + n_2} \xrightarrow[]{\nabla} \mathscr{P}(Q^{n_1+n_2}) \xrightarrow[]{\mathscr{P}(\Delta_\gamma \times \Delta_\varepsilon)} \mathscr{P}(\mathscr{P}(Q^{m_1}) \times \mathscr{P}(Q^{m_2})) \xrightarrow[]{\mathscr{P}(\nabla)} \mathscr{PP}(Q^{m_1 + m_2}) \xrightarrow[]{\mu} \mathscr{P}(Q^{m_1 + m_2}) \\
        &= \mathscr{P}(Q)^{n_1 + n_2} \xrightarrow[]{\nabla} \mathscr{P}(Q^{n_1+n_2}) \xrightarrow[]{\mathscr{P}(\Delta_\gamma \otimes \Delta_\varepsilon)} \mathscr{PP}(Q^{m_1 + m_2}) \xrightarrow[]{\mu} \mathscr{P}(Q^{m_1 + m_2}).
    \end{align*}

    Hence $\Delta_\gamma \otimes \Delta_\varepsilon$ is convex.
\end{proof}

\begin{proof}[Proof of Lemma \ref{lemma:detr}] \label{sec:lang_preserved}
  We need to show that this mapping is a morphism of pros. It is clear that identities are preserved. It remains to show that that composition and monoidal product are preserved. Let $\Delta_\alpha : Q^{a} \to \mathscr{P}(Q^b), \Delta_\beta : Q^b \to \mathscr{P}(Q^{c})$. We require $\detr_Q(\Delta_\beta \diamond \Delta_\alpha) = \detr_Q(\Delta_\beta) \circ \detr_Q(\Delta_\alpha)$. This follows from the commutativity of the following diagram (naturality of $\nabla$ and the naturality of $\eta$), and the unit law for Kleisli composition in $\Par$.
\[\begin{tikzcd}[ampersand replacement=\&]
	{(\bot\mathscr{P}^{+}(Q))^b} \& {(\bot\bot\mathscr{P}^{+}(Q))^b} \\
	{\bot\mathscr{P}^{+}(Q)^b} \& {\bot(\bot\mathscr{P}^{+}(Q))^b}
	\arrow["\nabla"', from=1-1, to=2-1]
	\arrow["{\bot\eta^b}"', from=2-1, to=2-2]
	\arrow["\eta", from=1-1, to=2-2]
	\arrow["\nabla", from=1-2, to=2-2]
	\arrow["{\eta^b}", from=1-1, to=1-2]
\end{tikzcd}\]

    Strict preservation of the monoidal product follows easily from the fact that $(\Delta_\gamma \otimes \Delta_\varepsilon)^{*} = \Delta_\gamma^{*} \times \Delta_\varepsilon^{*}$.
\end{proof}

\begin{proof}[Proof of Theorem \ref{thm:sufficient}] \label{appendix:sufficient}
We show that the defined mapping is indeed a morphism of pros. For composition, we need to show $\phi([\beta]\comp[\gamma]) = \phi([\beta])\comp\phi([\gamma])$. The $i^{\text{th}}$ component of $\phi([\beta]\comp[\gamma])([\alpha_1], ..., [\alpha_n])$ is the equivalence class:

\begin{center}
  \includegraphics[scale=0.7]{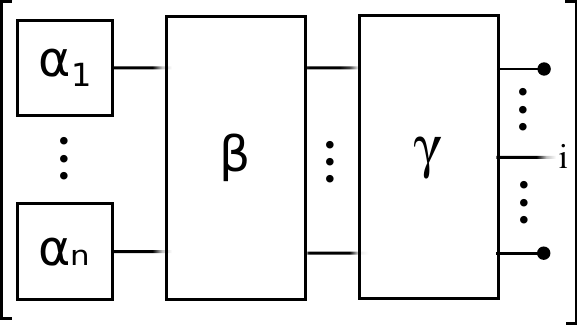}
\end{center}

where the $i^\text{th}$ output of $\gamma$ is dangling on the right. The $i^{\text{th}}$ component of $(\phi([\beta])\comp\phi([\gamma]))([\alpha_1], ..., [\alpha_n])$ is the equivalence class:

\begin{center}
  \includegraphics[scale=0.7]{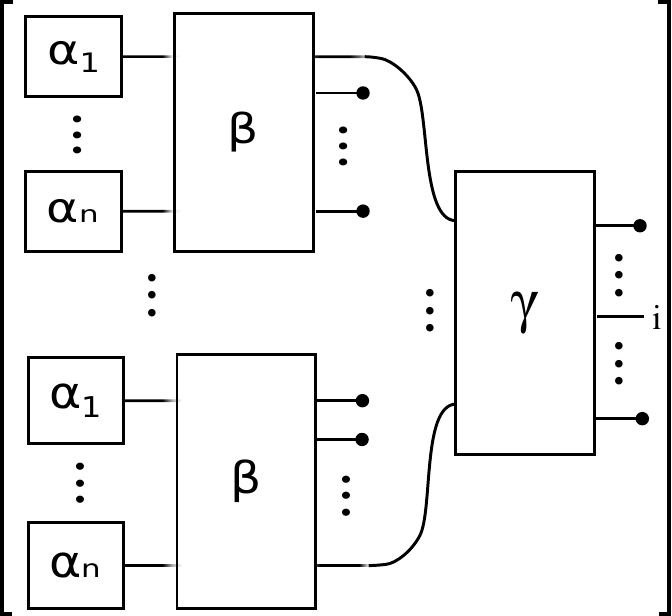}
\end{center}

But since $\mathscr{F}\Gamma/{\equiv_L}$ is a cartesian restriction prop, the representatives of these equivalence classes are the same diagram (by repeated applications of the naturality of \raisebox{.25em}{\begin{tikzpicture}[scale=.75,baseline=(current bounding box.center)]
\comult
\end{tikzpicture}
} and unitality). Hence this is the same equivalence class, and $\phi$ preserves composition.

For identities, the $i^{\text{th}}$ component of $\phi([\text{id}_n])([\alpha_1],...,[\alpha_n])$ is the equivalence class:

\begin{center}
  \includegraphics[scale=0.8]{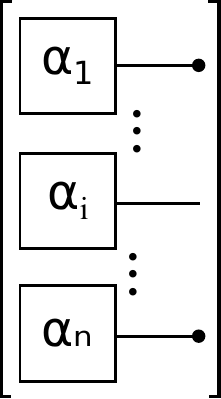}
\end{center}

For $\phi([\text{id}_n])$ to be the identity, this needs to be equal to the equivalence class $[\alpha_i]$:
\begin{center}
  \includegraphics[scale=0.8]{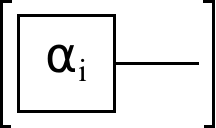}
\end{center}

But these must indeed be the same equivalence class, for if there were a context that distinguished these morphisms, we would have a contradiction, since languages are closed under monoidal products (Lemma \ref{lemma:closed_monoidal_product}). Similar diagrams hold for the preservation of the monoidal product, and thus we have a morphism of pros.
\end{proof}

\end{document}